\documentclass[11pt]{article}
\usepackage[margin=1in]{geometry}
\usepackage{amsmath,amssymb,amsthm,mathtools,hyperref,microtype}
\usepackage{tikz}
\usetikzlibrary{arrows.meta,calc}
\definecolor{walkblue}{RGB}{35,78,112}
\definecolor{walkrust}{RGB}{153,73,39}
\tikzset{
 manuscript diagram/.style={font=\small,>=Latex,line width=.7pt},
 diagram box/.style={draw=walkblue,fill=walkblue!4,rounded corners=2pt,
   align=center,inner sep=5pt},
 diagram arrow/.style={->,draw=walkblue,line width=.9pt},
 proof arrow/.style={->,draw=black!65,dashed},
 literal node/.style={circle,draw=walkblue,fill=white,
   minimum size=7mm,inner sep=1pt},
 sampled path/.style={->,draw=walkblue,line width=1pt},
 paired path/.style={->,draw=walkrust,dashed,line width=1pt}
}
\hypersetup{colorlinks=true,linkcolor=blue,citecolor=blue,urlcolor=blue,pdftitle={A One-Sided Sublinear-Query Tester for Bounded-Degree 2SAT}}
\newtheorem{theorem}{Theorem}[section]
\newtheorem{lemma}[theorem]{Lemma}

\newcommand{\E}{\mathbb E}
\newcommand{\ip}[2]{\langle #1,#2\rangle}
\newcommand{\pos}[1]{[#1]_+}
\newcommand{\Aff}{\operatorname{Aff}}
\newcommand{\KL}{\operatorname{D}}
\title{A One-Sided Sublinear-Query Tester for Bounded-Degree 2SAT}
\author{Yuichi Yoshida\thanks{Supported by JSPS KAKENHI Grant Numbers 22H05001 and 25K24465.}\\
National Institute of Informatics\\
\texttt{yyoshida@nii.ac.jp}
}
\RequirePackage[normalem]{ulem} 
\RequirePackage{color}\definecolor{RED}{rgb}{1,0,0}\definecolor{BLUE}{rgb}{0,0,1} 
\makeatletter 
\let\sout@orig\sout 
\renewcommand{\sout}[1]{\ifmmode\text{\sout@orig{\ensuremath{#1}}}\else\sout@orig{#1}\fi} 
\makeatother 
\RequirePackage{listings} 
\RequirePackage{color} 
\lstdefinelanguage{DIFcode}{ 
  moredelim=[il][\color{red}\sout]{\%DIF\ <\ }, 
  moredelim=[il][\color{blue}\uwave]{\%DIF\ >\ } 
} 
\lstdefinestyle{DIFverbatimstyle}{ 
	language=DIFcode, 
	basicstyle=\ttfamily, 
	columns=fullflexible, 
	keepspaces=true 
} 
\lstnewenvironment{DIFverbatim}{\lstset{style=DIFverbatimstyle}}{} 
\lstnewenvironment{DIFverbatim*}{\lstset{style=DIFverbatimstyle,showspaces=true}}{} 
\begin{document}
\maketitle
\begin{abstract}
We give a one-sided tester for satisfiability of 2CNF formulas in the bounded-degree model. For every degree bound $\Delta$, distance parameter $\varepsilon>0$, and $\nu\in(0,1/2)$, its query complexity on $n$-variable formulas is $\widetilde O_{\varepsilon,\Delta,\nu}(n^{1/2+\nu})$.
\end{abstract}
\thispagestyle{empty}
\newpage
\tableofcontents
\thispagestyle{empty}
\newpage
\setcounter{page}{1}

\section{Introduction}\label{sec:introduction}
We study whether satisfiability of bounded-degree 2CNF formulas can be tested with a sublinear number of queries. This is a question in \emph{property testing}, where the goal is to distinguish inputs having a property from those far from having it by inspecting only a small part of the input. For general background, we refer to the books by Goldreich~\cite{Gol17} and Bhattacharyya and Yoshida~\cite{BY22}.

As a basic example, in the bounded-degree model~\cite{GR02}, Goldreich and Ron gave a one-sided tester for bipartiteness with $\widetilde O(\sqrt n)$ queries and proved an $\Omega(\sqrt n)$ lower bound, where $n$ is the number of vertices~\cite{GR99,GR02}. Bipartiteness is a \emph{constraint satisfaction problem} (CSP): each vertex is a Boolean variable, and each edge requires its endpoints to take different values. More generally, a CSP specifies a finite domain of values and a finite collection of allowed constraint types. An instance consists of variables and constraints of these types, and asks whether an assignment satisfies all constraints. The degree of a variable is the number of constraints containing it. This setting includes 3LIN, whose constraints are linear equations modulo two on three variables. In contrast to bipartiteness, testing satisfiability of bounded-degree 3LIN requires $\Omega(n)$ queries, where $n$ is now the number of variables~\cite{BOT02}. The same linear lower bound holds for bounded-degree 3SAT, where clauses contain at most three literals~\cite{BOT02}.

A useful distinction between these problems comes from algorithms based on \emph{local consistency}. Such algorithms maintain candidate assignments to small sets of variables. They repeatedly discard candidates that violate an input constraint or cannot be extended consistently with the candidates on another small set. A CSP has \emph{bounded width} if this process decides satisfiability for all its instances while considering sets of at most a constant number of variables; the constant may depend on the allowed constraint types but not on the instance size~\cite{BKW17}. For example, in bipartiteness, $x\ne y$ and $y\ne z$ imply $x=z$. Propagating such relations between pairs detects the contradiction around any odd cycle. Thus bipartiteness has bounded width, even though the propagation may traverse the whole graph. Bounded width limits the number of variables considered together, not the amount of input read or the number of propagation steps.

At the simplest level, \emph{width~1}, it suffices to propagate the allowed values of individual variables through the constraints. It is known that CSPs of width~1 admit constant-query testers~\cite{Yos11,KOTYZ12}.
This propagation, however, does not suffice for bipartiteness. To see this, initialize each vertex of an odd cycle with both possible colors. Each color at either endpoint of an edge is compatible with the opposite color at the other endpoint, so this propagation removes no colors and fails to detect unsatisfiability.
Bipartiteness is a special case of 2SAT, since $x\ne y$ is equivalent to $(x\vee y)\wedge(\bar x\vee\bar y)$. Both problems have bounded width but not width~1: it suffices to maintain candidate assignments to pairs and repeatedly check that they extend consistently to each third variable. This is called \emph{width~$(2,3)$}~\cite{BKW17}.

In contrast, 3LIN has unbounded width. More generally, Fei proved an $\Omega(n)$ query lower bound for every unbounded-width CSP and asked whether every bounded-width CSP admits a sublinear-query tester~\cite[Question~12]{Fei26}. Beyond bipartiteness, 2SAT is a basic case of this question. It is known that testing 2SAT requires $\Omega(\sqrt n)$ queries~\cite{Yos11}, but the bipartiteness tester does not give an algorithm for general 2SAT. Fei and Rubinfeld also asked whether 2SAT admits a sublinear-query satisfiability tester~\cite[full version, Section~1.3]{FR27}.

In this paper, we answer this question affirmatively for 2SAT: For constant degree bound and distance parameter, we give a one-sided tester with $\widetilde O(n^{1/2+\nu})$ queries for every constant $\nu\in(0,1/2)$.

To state our main theorem precisely, we specify the input model and distance measure. 
The input $F$ is a 2CNF formula on $n\ge1$ Boolean variables labeled by $[n]\coloneqq \{1,\ldots,n\}$: a conjunction of nonempty clauses, each containing at most two literals.\footnote{Empty clauses cannot be detected through incidence queries.}
We identify $(a\vee a)$ with the unit clause $(a)$ and allow tautologies.
We assume throughout that no clause is repeated.\footnote{The tester also handles repeated clauses, with each occurrence counted separately in degrees, weighted sums, and the number of violated clauses. For this extension, each occurrence has a distinct identifier that is returned consistently at its incident variables.}
The size $n$ and the degree bound $\Delta$ are given to the tester. Each variable belongs to at most $\Delta$ clauses and has $\Delta$ incidence slots containing its incident clauses or null symbols. Unused slots may occur anywhere in the list. A query $(x,i)$, with $x\in[n]$ and $1\le i\le\Delta$, returns the clause in slot $i$, including its variables and literal signs, or the null symbol if that slot is unused. Let $\tau(F)$ denote the minimum number of clauses that must be deleted to make $F$ satisfiable, equivalently the minimum number of clauses violated by an assignment. We call $F$ \emph{$\varepsilon$-far} if $\tau(F)\ge\varepsilon\Delta n$. A \emph{one-sided $\varepsilon$-tester} accepts every satisfiable formula with probability one and rejects every $\varepsilon$-far formula with probability at least $2/3$.

Our main result is the following.
\begin{theorem}\label{thm:tester}
For every $\Delta\ge1$, $\varepsilon>0$, and $0<\nu<1/2$, there is a one-sided $\varepsilon$-tester for satisfiability of $n$-variable 2CNF formulas of maximum degree at most $\Delta$ in the bounded-degree model, with query complexity
\[
 \widetilde O_{\varepsilon,\Delta,\nu}(n^{1/2+\nu}).
\]
\end{theorem}
The $\Omega(\sqrt n)$ lower bound~\cite{Yos11} shows that the dependence on $n$ is optimal up to an arbitrarily small polynomial factor.

The picture changes for \emph{tolerant testing} of 2SAT, which also requires acceptance of formulas that are close to satisfiable. In fact, known lower bounds~\cite{Yos11CCC,CKKMP18} imply that, for some constants $0<\varepsilon_1<\varepsilon_2$, $\Delta\ge1$, and $c>0$, distinguishing $n$-variable 2CNF formulas of maximum degree at most $\Delta$ with $\tau(F)\le\varepsilon_1\Delta n$ from those with $\tau(F)\ge\varepsilon_2\Delta n$ requires $\Omega(n^{1/2+c})$ queries. Section~\ref{sec:related} gives the reduction to 2SAT. Thus, the guarantee for every fixed $0<\nu<1/2$ in Theorem~\ref{thm:tester} depends on requiring acceptance only of satisfiable formulas.

\subsection{Technical overview}\label{sec:overview}
We describe the tester for \emph{E2SAT}, where each clause contains literals on two distinct variables. A local reduction handles general 2SAT (Section~\ref{sec:general}). To analyze rejection on far inputs, we assume that every assignment violates at least a constant $\delta>0$ fraction of the clauses.

The \emph{implication graph} of $F$ has a vertex for each literal and arcs $\bar a\to b$ and $\bar b\to a$ for each clause $(a\vee b)$. Paths $s\leadsto\bar s$ and $\bar s\leadsto s$ rule out both truth values of $s$, so their clauses are unsatisfiable. Conversely, every unsatisfiable 2CNF formula contains such paths~\cite{APT79}. Our tester therefore searches for this certificate using only a small part of the formula.

\paragraph{Finding contradictions by endpoint collisions.}
Following~\cite{manuscript}, we search for contradictory paths by sampling forward walks. For every implication $u\to v$, the graph also contains $\bar v\to\bar u$. Thus, if two walks from $s$ end at complementary literals $z,\bar z$, reversing and complementing the second path gives $z\leadsto\bar s$. Joining it to the first path gives $s\leadsto\bar s$.

\begin{figure}[tbp]
 \centering
 \begin{tikzpicture}[manuscript diagram]
 \node[font=\small\bfseries] at (1.35,1.25) {Sampled forward paths};
 \node[font=\small\bfseries] at (10.2,1.25) {Recovered implication paths};

 \node[literal node] (s) at (0,0) {$s$};
 \node[literal node] (z) at (2.7,.55) {$z$};
 \node[literal node] (bz) at (2.7,-.55) {$\bar z$};
 \draw[sampled path] (s) -- (z);
 \draw[paired path] (s) -- (bz);
 \draw[->,draw=black!60] (4,0) -- (6.4,0)
   node[midway,above=3pt,align=center,font=\footnotesize]
   {reverse and\\complement};
 \node[literal node] (rs) at (7.4,0) {$s$};
 \node[literal node] (rz) at (10.2,0) {$z$};
 \node[literal node] (rbs) at (13,0) {$\bar s$};
 \draw[sampled path] (rs) -- (rz);
 \draw[paired path] (rz) -- (rbs);

 \node[literal node] (bs) at (0,-2.15) {$\bar s$};
 \node[literal node] (y) at (2.7,-1.6) {$u$};
 \node[literal node] (by) at (2.7,-2.7) {$\bar u$};
 \draw[sampled path] (bs) -- (y);
 \draw[paired path] (bs) -- (by);
 \draw[->,draw=black!60] (4,-2.15) -- (6.4,-2.15)
   node[midway,above=3pt,align=center,font=\footnotesize]
   {reverse and\\complement};
 \node[literal node] (rrbs) at (7.4,-2.15) {$\bar s$};
 \node[literal node] (ry) at (10.2,-2.15) {$u$};
 \node[literal node] (rrs) at (13,-2.15) {$s$};
 \draw[sampled path] (rrbs) -- (ry);
 \draw[paired path] (ry) -- (rrs);
\end{tikzpicture}

 \caption{Recovering a contradiction from four forward paths. In the first row,
 two paths from $s$ end at complementary literals $z,\bar z$; in the second,
 two paths from $\bar s$ end at complementary literals $u,\bar u$.
 Reversing and complementing each dashed path gives the dashed segment on
 the right. Every arrow between literal nodes denotes an implication path.
 Together, the two rows give $s\leadsto\bar s\leadsto s$,
 using only clauses on the sampled paths.}
 \label{fig:collision-certificate}
\end{figure}
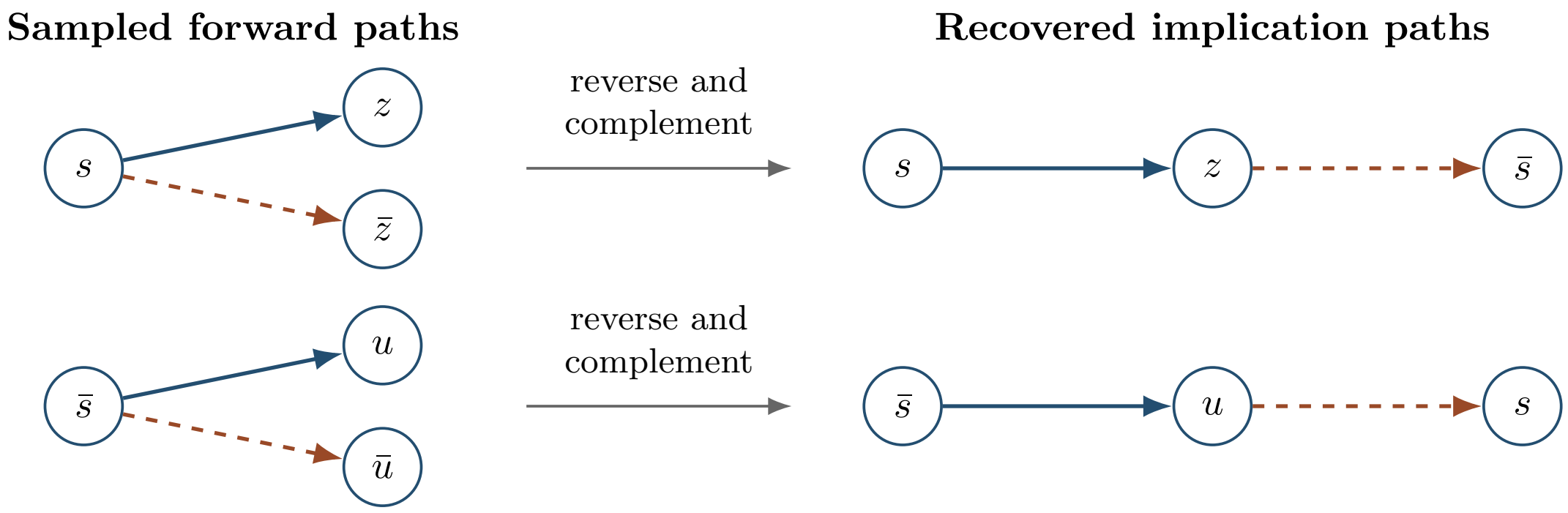

\paragraph{The tester.}
The tester uses clause weights $w_e\in(0,1]$, whose construction is outlined later in this overview. It repeats the following trial a prescribed number of times.
\begin{enumerate}
\item \emph{Sample a seed.} Choose a clause with probability proportional to its weight, choose one of its two literals uniformly, and take its complement as the seed $s$.
\item \emph{Explore from both $s$ and $\bar s$.} Run four independent batches of lazy forward walks, two from each starting literal, with independent geometric stopping and expected length $O(\log n)$. For a non-lazy step from $\ell$, select a clause containing $\bar\ell$ with probability proportional to its weight and follow its implication to the other literal. Retain the paths as well as their endpoints.
\item \emph{Look for a contradiction.} In the two batches from $s$, look for an endpoint in one batch whose complement occurs in the other. Perform the same check for the two batches from $\bar s$. If both checks succeed, reject and output the paths $s\leadsto\bar s\leadsto s$ obtained by reversing and complementing paths as above (Figure~\ref{fig:collision-certificate}). Otherwise try a new seed.
\end{enumerate}
If all trials finish without a contradiction, accept. Every rejection produces an unsatisfiable set of input clauses, so satisfiable inputs are always accepted. The main issue is to choose locally computable weights that make these trials succeed on far inputs within the query budget.

\paragraph{Measuring the chance of finding a contradiction.}
Let $\mathcal L$ be the literal set. For each $s\in\mathcal L$, let $p_s:\mathcal L\to[0,1]$ be the endpoint probability mass function of a walk from $s$. Define the \emph{affinity function} $a:\mathcal L\to[0,1]$ by
\[
 a(s)\coloneqq\sum_{\ell\in\mathcal L}\sqrt{p_s(\ell)p_s(\bar\ell)}.
\]
Thus $a(s)$ is the \emph{Hellinger affinity} between $p_s$ and its complemented distribution. The collision bound of Yoshida~\cite[Lemma~2.1]{manuscript} implies that constant $a(s)$ suffices to find complementary endpoints with $O(\sqrt n)$ walks; the cost for smaller $a(s)$ grows polynomially in $1/a(s)$. For completeness, we prove this bound in Lemma~\ref{lem:collision}.
However, to apply this collision bound in both directions for the same seed, we seek lower bounds on both $a(s)$ and $a(\bar s)$. We therefore define the \emph{two-way affinity} $B\in[0,1]$ by
\[
 B\coloneqq\E_s\min\{a(s),a(\bar s)\},
\]
where the expectation is over the seed distribution. The minimum makes each seed contribute only as much as its weaker direction allows. Our task is to choose walks and seeds so that $B$ is large on every far input, while keeping the sampling cost small.

\paragraph{Choosing weights that can be used locally.}
Let $m$ be the number of clauses. We use clause weights to define both the walks and the seed distribution. Write a clause $(\bar u\vee\bar v)$ as a pair $e=\{u,v\}\subseteq\mathcal L$, with endpoints $u,v$; it is violated exactly when both endpoints are true. For a weight vector $w\in\mathbb R_{\ge0}^m$ with positive total weight, let $\pi\coloneqq w/\sum_e w_e\in[0,1]^m$ be the corresponding clause probability distribution. Sampling a clause from $\pi$ and choosing one of its endpoints uniformly gives the literal distribution $\mu_\pi:\mathcal L\to[0,1]$ defined by
\[
 \mu_\pi(\ell)\coloneqq\frac{1}{2}\sum_{e\ni\ell}\pi_e.
\]
Thus the seed distribution is $\mu_\pi$, which need not be stationary. The walk uses the same weights for its implication transitions.

The analysis of~\cite{manuscript} uses weights for which $\mu_\pi$ is stationary and gives equal mass to complementary literals. We call the weights \emph{balanced} if $\mu_\pi(\ell)=\mu_\pi(\bar\ell)$ for every literal $\ell$. One implication step from $\mu_\pi$ ends at $\ell$ with probability $\mu_\pi(\bar\ell)$. Thus balance makes $\mu_\pi$ stationary for both the implication walk and its lazy version.

Uniform weights need not be balanced. Changing weights to improve balance, however, could make the clauses responsible for unsatisfiability carry too little weight. Moreover, computing the weights by reading the entire formula would defeat the purpose of a sublinear-query tester. We therefore seek locally computable weights with small imbalance and a positive lower bound on the fraction of total weight carried by clauses violated by any assignment.

We obtain these properties by choosing a family of clause-weight vectors $w(z)\in(0,1]^m$, parameterized by $z\in\mathbb R^n$, together with a convex objective $\Phi:\mathbb R^n\to\mathbb R_{\ge0}$. The coordinates $z_x$ are indexed by variables, and the weights $w_e(z)$ are indexed by clauses. To express imbalance in terms of clause weights, define the \emph{weighted degree} $d_w:\mathcal L\to\mathbb R_{\ge0}$ by $d_w(\ell)\coloneqq\sum_{e\ni\ell}w_e$. Since $\mu_\pi(\ell)=d_w(\ell)/(2\sum_e w_e)$, balance is equivalent to $d_w(x)=d_w(\bar x)$ for every variable $x$. We choose $w$ and $\Phi$ so that
\[
 \frac{\partial\Phi}{\partial z_x}(z)
 =-\bigl(d_{w(z)}(x)-d_{w(z)}(\bar x)\bigr).
\]
Thus small gradients mean small imbalance. Starting from $z=0$, which gives uniform weights, gradient descent keeps $\Phi(z)\le\Phi(0)=m$. We design $\Phi$ and $w(z)$ so that this inequality also prevents any assignment from violating clauses of too little total weight (Lemma~\ref{lem:sublevel}).

To see the idea, extend $z$ to literals by $z_{\bar x}\coloneqq-z_x$. We choose the weights so that a clause $e=\{u,v\}$ has very small weight only when $z_u+z_v$ is large and positive. Such a clause has an endpoint $\ell$ with large positive $z_\ell$. Making this endpoint false satisfies the clause. Given an assignment with small weighted loss, we therefore reassign variables with large $|z_x|$ in this way. A random threshold determines which variables to reassign: clauses on unchanged variables have weights bounded below, while $\Phi(z)\le m$ controls the expected number of violations involving reassigned variables. If the original weighted loss were sufficiently small, some choice of threshold would yield an assignment violating fewer than $\delta m$ clauses, contradicting the assumption on $F$.

Descent controls the imbalance averaged over iterations, but selecting a good iterate by measuring its total imbalance would require global information. Instead, the tester uses the average of the clause weights $w_e(z)$ over the iterates. For each assignment, the total weight of violated clauses is linear in the clause weights, so averaging preserves its lower bound. At the same time, averaging makes the \emph{relative imbalance},
\[
 \frac{\sum_x|d_w(x)-d_w(\bar x)|}{\sum_e w_e},
\]
arbitrarily small.
For any fixed imbalance tolerance, the required number of steps is independent of $n$. Each update uses only variables sharing a clause, so each final clause weight can be recovered by exploring a neighborhood of bounded radius (Theorem~\ref{thm:local}). The construction also guarantees $\sum_e w_e=\Omega(m)$. Together with local evaluation, this ensures that sampling a seed or taking a walk step uses an expected number of queries independent of $n$ on far inputs, without computing the total weight $\sum_e w_e$.

\paragraph{Using exact balance in the analysis.}
The locally computed weights are only approximately balanced, so their seed distribution $\mu_\pi$ may still fail to be stationary. The entropy argument below uses stationarity to average changes along walk transitions. We therefore construct an exactly balanced clause distribution $\widehat{\pi}\in[0,1]^m$ for the analysis, while the tester continues to use $\pi$ for clause choices and $\mu_\pi$ for seeds.

For a common stopping probability, write $B_\pi$ and $B_{\widehat{\pi}}$ for the two-way affinities of walks defined by $\pi$ and $\widehat{\pi}$, with seeds drawn from $\mu_\pi$ and $\mu_{\widehat{\pi}}$, respectively. We will prove a constant lower bound on $B_{\widehat{\pi}}$ and transfer it to $B_\pi$. This requires controlling how the seed and walk distributions change. The seed distribution is used once, whereas differences in the conditional clause distributions accumulate along the walk. We therefore need tighter control of the latter.

To obtain $\widehat{\pi}$, we reduce clause weights subject to exact balance. We maximize the minimum total weight of violated clauses over all assignments, minus a penalty given by the weighted relative entropies between the conditional clause distributions. Theorem~\ref{thm:comparison} shows that, when the initial imbalance is sufficiently small, an optimizer retains a positive lower bound on the total weight of clauses violated by any assignment. We normalize these weights to obtain $\widehat{\pi}$. By adjusting the penalty and reducing the initial imbalance, we can make the average relative entropy between the conditional clause distributions arbitrarily small, while preserving a positive lower bound on the total weight of violated clauses under $\widehat{\pi}$ independently of this tolerance. The average is over literals drawn from $\mu_{\widehat{\pi}}$. The relative entropy between the seed distributions remains bounded by a constant. Since $\widehat{\pi}$ is used only in the proof, its construction need not be local.

\paragraph{Finding paths from $s$ to $\bar s$ and back.}
We now consider walks defined by the balanced distribution $\widehat{\pi}$ and write $a(s)$ for their endpoint affinity. The entropy and rounding argument of~\cite{manuscript} gives a positive constant lower bound on $\E_{s\sim\mu_{\widehat{\pi}}}a(s)$ for walks of logarithmic expected length. The argument proceeds by contradiction: if this average affinity were small, Gaussian rounding of vectors constructed from the endpoint distributions would give an assignment whose violated clauses have small total weight, contradicting the guarantee for $\widehat{\pi}$ (Section~\ref{sec:average-affinity}).

This still leaves a difficulty: a large average of $a(s)$ does not ensure that $a(s)$ and $a(\bar s)$ are large at the same seeds. To lower-bound $B_{\widehat{\pi}}$, we seek a set of variables on which $a(s)$ and $a(\bar s)$ are close and every assignment violates internal clauses of total weight bounded below by a positive constant. A lower bound on this set's contribution to the average affinity will then give a lower bound on the average of $\min\{a(s),a(\bar s)\}$.

To find such a set, we assign variables $x$ for which $a(x)$ and $a(\bar x)$ differ substantially, making their literal with smaller affinity true. The key estimate is that the difference $a(\bar s)-a(s)$ changes little on average along implications (Lemma~\ref{lem:stability}). A clause violated by this partial assignment forces a large change in that difference, so the total weight of clauses violated by this partial assignment is small. Choosing the assignment threshold by averaging also keeps small the weight of clauses crossing between assigned and unassigned variables (Lemma~\ref{lem:threshold}). Consequently, every assignment to the remaining variables must still violate internal clauses of total weight bounded below by a positive constant, and their complementary literals have close affinities.

\paragraph{Restricting the walk to the remaining variables.}
Let $S\subseteq\mathcal L$ be the set of literals of the remaining variables. Simply deleting the other literals and their incident clauses could destroy balance, so the walk on the internal clauses need not have a stationary distribution that is equal at complementary literals. Instead, whenever the walk leaves $S$, we follow it until it returns and treat the whole excursion as one step. This \emph{trace chain} has stationary distribution $\mu_{\widehat{\pi}}$ conditioned on $S$, which remains equal at complementary literals. It also retains every original transition within $S$ (Lemma~\ref{lem:trace}). We prove the average-affinity bound for any lazy chain with these two properties, so the lower bound on the total weight of internal clauses violated by any assignment gives the bound directly for the trace chain (Lemma~\ref{lem:oneway}).

We must still relate its endpoints to those of the original walk: a single trace step can take many original steps, during which the original walk might stop. The probability $\mu_{\widehat{\pi}}(S)$ that the seed lies in $S$ is bounded below by a positive constant, which bounds the expected return time when starting from the conditional stationary distribution. By choosing the original walk's stopping probability to be a sufficiently small constant multiple of the trace chain's stopping probability, we control the mass lost to stopping before a return and transfer a positive affinity bound back to the original walks from $S$ (Lemma~\ref{lem:comparison}). Since $a(s)$ and $a(\bar s)$ are close for $s\in S$, this yields the desired constant lower bound on $B_{\widehat{\pi}}$ (Theorem~\ref{thm:geo}).

\paragraph{From the analysis to the query bound.}
Finally, we transfer the guarantee to the tester's walks. If the average relative entropy between the conditional clause distributions is at most $\alpha>0$, its accumulation over logarithmically many expected steps gives a divergence bound of $O_\delta(1+\alpha\log n)$ for the joint seed and endpoint distributions (Lemma~\ref{lem:walkkl}). The resulting lower bound on $B_\pi$ decreases exponentially with this divergence, so the loss in two-way affinity is only a factor of $n^{O_\delta(\alpha)}$ (Lemma~\ref{lem:kltransfer}).

\begin{figure}[tbp]
 \centering
 \begin{tikzpicture}[manuscript diagram,
 overview box/.style={diagram box,text width=2.9cm,minimum height=1.12cm}]
 \node[anchor=west,font=\small\bfseries] at (-1.65,1.02) {Tester};
 \node[overview box] (weights) at (0,0)
   {Locally computed\\weights $w$\\Section~\ref{sec:localweights}};
 \node[overview box] (walks) at (4,0)
   {Forward walks\\sampled with $w$};
 \node[overview box] (tests) at (8,0)
   {Both endpoint tests\\from $s$ and $\bar s$};
 \node[overview box] (certificate) at (12,0)
   {Certificate\\$s\leadsto\bar s\leadsto s$};
 \draw[diagram arrow] (weights) -- (walks);
 \draw[diagram arrow] (walks) -- (tests);
 \draw[diagram arrow] (tests) -- (certificate)
   node[midway,above=24pt,font=\scriptsize] {if both tests succeed};

 \node[anchor=west,font=\small\bfseries] at (2.35,-1.5)
   {Analysis on far inputs};
 \node[overview box,draw=black!60,fill=black!2] (pi) at (0,-2.85)
   {Approximately\\balanced $\pi$\\Every assignment\\violates clauses of\\total weight $\Omega(1)$};
 \node[overview box,draw=black!60,fill=black!2] (rho) at (4,-2.85)
   {Balanced $\widehat{\pi}$\\Section~\ref{sec:comparison}};
 \node[overview box,draw=black!60,fill=black!2] (rhooverlap) at (8,-2.85)
   {Two-way affinity $B_{\widehat{\pi}}$\\Section~\ref{sec:geo}};
 \node[overview box,draw=black!60,fill=black!2] (pioverlap) at (12,-2.85)
   {Two-way affinity $B_\pi$\\Section~\ref{sec:transfer}};
 \draw[proof arrow] (weights.south) -- (pi.north)
   node[midway,left=4pt,font=\footnotesize] {$\pi=\dfrac{w}{\sum_e w_e}$};
 \draw[proof arrow] (pi) -- (rho);
 \draw[proof arrow] (rho) -- (rhooverlap);
 \draw[proof arrow] (rhooverlap) -- (pioverlap);
 \draw[proof arrow] (pioverlap.north) -- (12,-1.15) -- (8,-1.15)
   node[midway,below=3pt,font=\footnotesize]
   {collisions (Section~\ref{sec:collision})}
   -- (tests.south);
\end{tikzpicture}

 \caption{The tester and its analysis. The upper row
 shows the operations performed using locally evaluated weights $w$.
 The lower row records the argument on far inputs: normalize to $\pi$,
 establish the existence of balanced
 $\widehat{\pi}$, lower-bound its two-way affinity $B_{\widehat{\pi}}$, and transfer
 the bound to $B_\pi$. The two endpoint tests must both
 succeed for the same seed: one uses walks from $s$, the other from $\bar s$.
 Dashed arrows indicate logical implications in the analysis. The collision
 lemma gives a lower bound on the probability that both endpoint tests succeed.}
 \label{fig:tester-overview}
\end{figure}
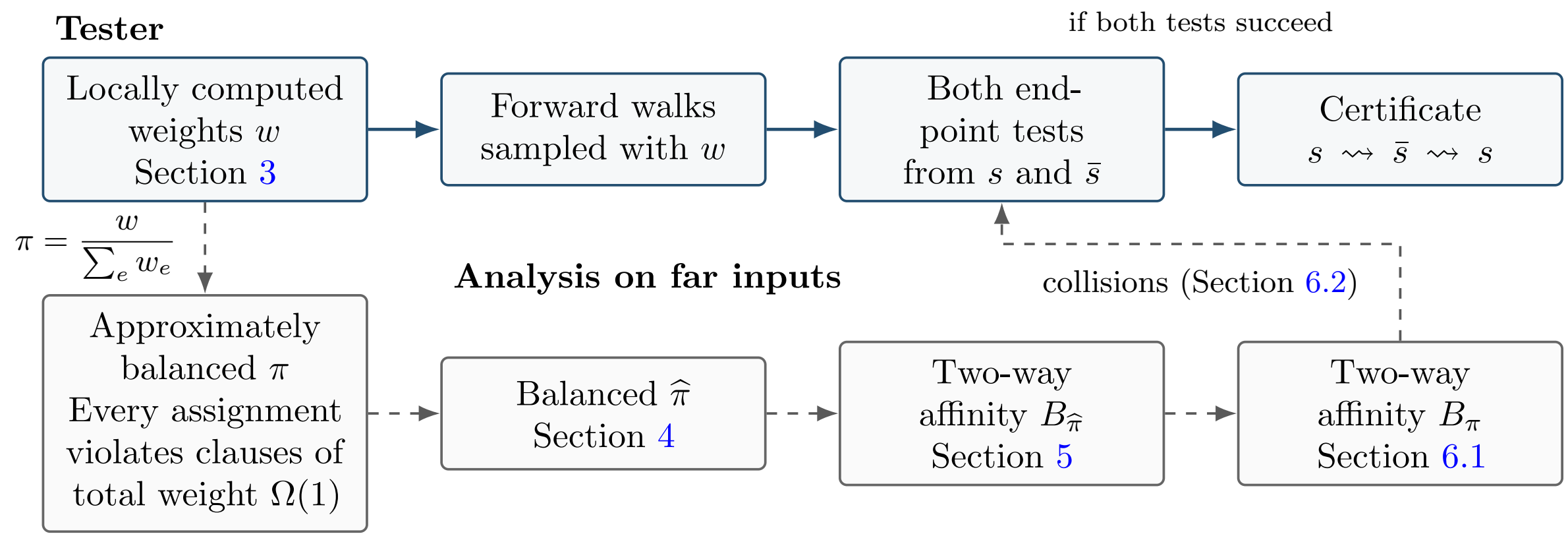

The collision search and the number of seeds tried incur only polynomial costs in $1/B_\pi$. Thus choosing $\alpha$ sufficiently small makes their combined overhead at most $n^\nu$ on top of the $\sqrt n$ collision bound. Each walk has logarithmic expected length, and each weight evaluation has cost independent of $n$, giving the $\widetilde O_{\varepsilon,\Delta,\nu}(n^{1/2+\nu})$ query bound. Section~\ref{sec:algorithm} specifies the parameters and implements the sampling within finite query budgets. Figure~\ref{fig:tester-overview} summarizes the tester and its analysis.

\subsection{Related work}\label{sec:related}
Krom gave the first polynomial-time algorithm for 2SAT~\cite{Krom67}, and Aspvall, Plass, and Tarjan gave a linear-time algorithm based on strongly connected components of the implication graph~\cite{APT79}. In the sublinear-query setting, Fei and Rubinfeld~\cite{FR27} gave a new analysis of the bipartiteness tester of Goldreich and Ron~\cite{GR99}, reducing the number of walks from $O_\varepsilon(\sqrt n\log n)$ to $O_\varepsilon(\sqrt n)$ and their length from $O_\varepsilon(\log^6 n)$ to $O_\varepsilon(\log n)$. This yields a query complexity of $O_\varepsilon(\sqrt n\log n)$. They also ask whether, for some constant $0<\varepsilon_0<1/4$, a sublinear-query tester can distinguish satisfiable bounded-degree 2CNF formulas from formulas in which every assignment satisfies at most $3/4+\varepsilon_0$ of the clauses~\cite[full version, Problem~1.12]{FR27}.

Yoshida's unpublished manuscript~\cite{manuscript} gives a tester for E2SAT under this promise, assuming access to a uniform clause sampler in addition to incidence queries. Our tester builds on its random-walk analysis to handle every constant distance from satisfiability using only incidence queries. In particular, on instances with $\Theta(n)$ clauses, Theorem~\ref{thm:tester} distinguishes satisfiable formulas from formulas in which every assignment satisfies at most a $1-\delta$ fraction of the clauses, for every fixed $\delta>0$. Taking $\delta=1/4-\varepsilon_0$ gives the promise in Fei and Rubinfeld's question.
Section~\ref{sec:overview} explains which parts of the manuscript's analysis we use. All ingredients used from that manuscript are proved here, so the reader need not consult it.

For general bounded-degree CSPs, Yoshida~\cite{Yos11} gives constant-time approximation algorithms whose guarantees are governed by the integrality gap of the basic LP relaxation. For satisfiability testing, this approach yields constant-query testers for width-1 CSPs~\cite{Yos11,KOTYZ12}, but does not cover all bounded-width CSPs. The obstruction is that even for 2SAT, the basic LP can have value $1$ on instances that are a constant distance from satisfiability~\cite{KOTYZ12}.

The tolerant-testing lower bound for 2SAT discussed after Theorem~\ref{thm:tester} follows from lower bounds for 2-XOR and Max Cut. Yoshida~\cite[Theorem~1.2]{Yos11CCC} proves such a bound for 2-XOR, whose constraints have the form $x\oplus y=b$ for Boolean variables $x,y$ and $b\in\{0,1\}$. For every fixed $0<\eta<1/4$, there are constants $\Delta=O(\eta^{-2})$ and $\delta=\delta(\eta)>0$ such that distinguishing $n$-variable instances of maximum degree at most $\Delta$ that can be made satisfiable by deleting at most $\eta\Delta n/2$ constraints from those requiring at least $(1/2-\eta)\Delta n/2$ deletions requires $\Omega(n^{1/2+\delta})$ queries.

For Max Cut, Chiplunkar, Kapralov, Khanna, Mousavifar, and Peres~\cite[Theorem~6]{CKKMP18} show that estimating the maximum cut size within a factor $2-\rho$, for fixed $0<\rho<1/2$, requires $n^{1/2+\Omega(\rho/\log(1/\rho))}$ queries, even on graphs of maximum degree $O(\rho^{-2})$. They attribute this Max Cut lower bound to Yoshida~\cite{Yos11CCC} and give a different proof~\cite[full version, Remark~1]{CKKMP18}. Their construction separates graphs having a cut that contains a $1-O(\rho)$ fraction of the edges from graphs in which every cut contains at most a $1/2+O(\rho)$ fraction. The former graphs need not be bipartite.

The Max Cut lower bound transfers to tolerant testing of 2SAT by a local encoding. For a graph $G=(V,E)$, form a 2CNF formula $F_G$ on the Boolean variables $\{x_v:v\in V\}$ by replacing each edge $\{u,v\}\in E$ with $(x_u\vee x_v)\wedge(\bar x_u\vee\bar x_v)$. Write $\operatorname{MaxCut}(G)$ for the maximum number of edges crossing a cut of $G$. Each edge outside a cut violates exactly one of its two clauses, while each crossing edge satisfies both, so
\[
 \tau(F_G)=|E|-\operatorname{MaxCut}(G).
\]
The encoding doubles the degree bound and simulates each formula query with a constant number of graph queries. Moreover, for each fixed sufficiently small $\rho$, the graphs used in the lower bound have $\Theta(n)$ edges. Hence the resulting formulas satisfy $\tau(F_G)\le\varepsilon_1\Delta n$ in the first case and $\tau(F_G)\ge\varepsilon_2\Delta n$ in the second, for some constants $0<\varepsilon_1<\varepsilon_2$, where $\Delta$ is the formula degree bound. This gives the tolerant-testing lower bound stated after Theorem~\ref{thm:tester}.

In the dense model, constant-query testers are known for satisfiability of CSP instances \cite{AE02,AS03}. Those results use a dense representation and normalize distance by the number of possible variable tuples for constraints; their guarantees therefore do not directly apply in the bounded-degree model considered here.

\paragraph{Organization.}
Section~\ref{sec:preliminaries} introduces the common notation and proves a threshold assignment lemma. Section~\ref{sec:localweights} constructs the locally computable clause weights used by the tester. Section~\ref{sec:comparison} provides an exactly balanced clause distribution for comparison, with bounds on how the seed and conditional clause distributions change. Section~\ref{sec:geo} proves a lower bound on the two-way affinity for walks defined by this balanced distribution. Finally, Section~\ref{sec:tester} transfers the bound to the tester's walks, uses endpoint collisions to detect contradictions, and proves the query bound, including the reduction from general 2SAT to E2SAT.

\section{Preliminaries}\label{sec:preliminaries}
We fix notation for weighted E2SAT formulas and probability measures and record a threshold assignment lemma. Write $\pos{c}\coloneqq\max\{c,0\}$ for $c\in\mathbb R$.

Formulas with no clauses or maximum degree at most one are satisfiable, and $\varepsilon>1$ admits no far inputs. We therefore restrict the analysis to formulas with at least one clause, degree bounds at least two, and $0<\varepsilon\le1$.

\subsection{Weighted E2SAT formulas}\label{sec:weighted-notation}
Let $F$ be an E2SAT formula with $m$ clauses on $n$ variables, so each clause contains literals on two distinct variables. Write $\mathcal L\coloneqq \{x,\bar x:x\text{ is a variable of }F\}$ for its literals. We represent a clause $(\bar u\vee\bar v)$ by the pair $e=\{u,v\}\subseteq\mathcal L$. The clause is violated exactly when both endpoints $u$ and $v$ are true. We index variable vectors by $x$ and clause vectors by $e$, writing their coordinates as $z_x$ and $w_e$, respectively. Let $\mathcal T\coloneqq \{t\in\{0,1\}^{\mathcal L}:t_{\bar x}=1-t_x\text{ for every variable }x\}$ be the assignment space. For a weight vector $w\in\mathbb R_{\ge0}^m$, define its total weight $W(w)\in\mathbb R_{\ge0}$, its \emph{weighted loss function} $L_w:\mathcal T\to\mathbb R_{\ge0}$, and its \emph{minimum weighted loss} $\tau_w(F)\in\mathbb R_{\ge0}$ by
\begin{equation}\label{eq:weightedloss}
 W(w)\coloneqq \sum_e w_e,\qquad L_w(t)\coloneqq \sum_{e=\{u,v\}}w_et_ut_v,
 \qquad \tau_w(F)\coloneqq \min_{t\in\mathcal T} L_w(t).
\end{equation}
We omit zero-weight clauses from weighted sums. Assignments specified on all variables incident to positive-weight clauses are extended arbitrarily to the remaining variables.

For $W(w)>0$, the ratio $\tau_w(F)/W(w)$ is the minimum, over all assignments, of the total weight of violated clauses divided by the total clause weight. To measure the difference between the incident weights at complementary literals, define the weighted degree $d_w:\mathcal L\to\mathbb R_{\ge0}$, the \emph{imbalance} $b:\mathbb R_{\ge0}^m\to\mathbb R^n$, and the \emph{total imbalance} $\|b(w)\|_1$ by
\begin{equation}\label{eq:imbalance}
 d_w(\ell)\coloneqq \sum_{e\ni\ell}w_e,\qquad
 b_x(w)\coloneqq d_w(x)-d_w(\bar x),\qquad \|b(w)\|_1=\sum_x|b_x(w)|.
\end{equation}
Here $b_x(w)$ is the coordinate of $b(w)$ at variable $x$, and $w$ is balanced precisely when $b(w)=0$. For a clause probability distribution $\sigma\in[0,1]^m$, the induced literal distribution $\mu_\sigma:\mathcal L\to[0,1]$ is given by $\mu_\sigma(\ell)\coloneqq \tfrac12\sum_{e\ni\ell}\sigma_e$. This is the distribution obtained by drawing a clause from $\sigma$ and choosing one of its endpoints uniformly. Thus $\mu_{w/W(w)}(\ell)=d_w(\ell)/(2W(w))$ whenever $W(w)>0$. A clause distribution $\sigma$ is balanced precisely when $\mu_\sigma(x)=\mu_\sigma(\bar x)$ for every variable $x$.

For $\mu_\sigma(\ell)>0$, let $q_\sigma(\cdot\mid\ell)$ be the probability distribution on clauses incident to $\ell$ given by
\[
 q_\sigma(e\mid\ell)\coloneqq\frac{\sigma_e}{2\mu_\sigma(\ell)}
 \qquad(e\ni\ell).
\]
This is the conditional law of the sampled clause given that its selected endpoint is $\ell$.

\subsection{Probability notation}\label{sec:probability-notation}
We use relative entropy to compare sampling distributions and Hellinger affinity to measure their overlap. Throughout, $\log$ denotes the natural logarithm, and $\log_2$ denotes the logarithm to base two.

For probability mass functions $p,q:\Omega\to[0,1]$ on a finite set $\Omega$, define their \emph{relative entropy} by $\KL(p\|q)\coloneqq \sum_{i\in\Omega}p_i\log(p_i/q_i)\in[0,+\infty]$, with $p_i\coloneqq p(i)$, $q_i\coloneqq q(i)$, zero summands interpreted as zero and divergence $+\infty$ if $p_i>0=q_i$ for some $i$.

For any finite set $\Omega$, define the Hellinger affinity $\Aff:\mathbb R_{\ge0}^{\Omega}\times\mathbb R_{\ge0}^{\Omega}\to\mathbb R_{\ge0}$ of nonnegative mass functions $p,q$ by
\[
 \Aff(p,q)\coloneqq \sum_{z\in\Omega}\sqrt{p(z)q(z)}.
\]
For probability measures it lies in $[0,1]$; we also use the definition for \emph{subprobability measures}, whose total mass is at most one.

\subsection{Random threshold assignments}\label{sec:threshold-assignment}
The next lemma bounds violations involving variables assigned by a random threshold, regardless of how the remaining variables are filled in. Its second part also counts clauses with one assigned and one unassigned variable before completion. We use these bounds in Sections~\ref{sec:localweights}, \ref{sec:comparison}, and~\ref{sec:geo}.

\begin{lemma}[Assignment using a random threshold]\label{lem:randomthreshold}
For an E2SAT formula with literal set $\mathcal L$, let $q\in\mathbb R^{\mathcal L}$ satisfy $q_{\bar\ell}=-q_\ell$, and let $0\le\alpha<\beta$. Choose $\theta$ uniformly in $[\alpha,\beta]$. For each variable $x$ with $|q_x|>\theta$, make its literal $\ell$ with $q_\ell>0$ true; leave the other variables unassigned. For every clause $e=\{u,v\}$, the following bounds hold.
\begin{enumerate}
\item[(i)] For any rule for completing this partial assignment, the probability that $e$ is violated and at least one endpoint was assigned by the threshold is at most $\pos{q_u+q_v}/(\beta-\alpha)$.
\item[(ii)] Before completion, the probability that exactly one endpoint is assigned, or both endpoints are assigned and $e$ is violated, is at most $|q_u+q_v|/(\beta-\alpha)$.
\end{enumerate}
\end{lemma}
\begin{proof}
For (i), order the endpoints so that $q_u\ge q_v$. If $e$ is violated and at least one endpoint was assigned by the threshold, then $q_u>\theta$ and $q_v\ge-\theta$. Thus $\theta$ belongs to $[-q_v,q_u)\cap[\alpha,\beta]$, an interval of length at most $\pos{q_u+q_v}$. Dividing by $\beta-\alpha$ gives (i), regardless of how the unassigned variables are filled in.

For (ii), if $q_u,q_v>0$, the event occurs exactly when $\theta<\max\{q_u,q_v\}$. Since $\theta\ge0$, the corresponding interval has length at most $\max\{q_u,q_v\}\le q_u+q_v$. Otherwise, $e$ cannot be violated when both endpoints are assigned. Exactly one endpoint is assigned only for thresholds between $|q_u|$ and $|q_v|$, an interval of length $\bigl||q_u|-|q_v|\bigr|\le|q_u+q_v|$. Dividing by $\beta-\alpha$ gives (ii).
\end{proof}

\section{Locally Computable Clause Weights}\label{sec:localweights}
Our goal is to construct the clause weights $w$ that the tester will use to sample seeds and walks, with each weight computable from a bounded neighborhood. The weights must have small relative imbalance while ensuring that, for every assignment to a far input, the violated clauses carry a positive fraction of the total weight. Theorem~\ref{thm:local} gives these guarantees; Section~\ref{sec:comparison} will use them to obtain an exactly balanced distribution for the analysis. We first design the convex objective $\Phi$ introduced in Section~\ref{sec:overview}, whose decrease preserves a lower bound on $\tau_w(F)$, then use gradient descent to reduce imbalance and show how to evaluate the resulting weights locally.

Throughout this section, $F$ is an E2SAT formula with $m$ clauses, maximum variable degree $\Delta$, and $\tau(F)\ge\delta m>0$.

\subsection{A lower bound on the minimum weighted loss}
We first construct a convex function $\Phi$ such that $\Phi(z)\le m$ ensures a lower bound on the minimum weighted loss $\tau_{w(z)}(F)$ for the weights $w(z)$ defined below. This will let us reduce the imbalance $b(w(z))$ by gradient descent while maintaining this lower bound. To make the gradient equal the negative imbalance, we will take each clause weight to be the negative derivative of its contribution to $\Phi$. An exponential contribution gives positive, decreasing weights, but its derivative is unbounded on negative arguments. Continuing it linearly there caps the weights at one and gives the globally Lipschitz derivative needed for fixed-step gradient descent.

Take a vector $z\in\mathbb R^n$ and extend it to literals by $z_{\bar x}\coloneqq -z_x$. Define the linear map $c:\mathbb R^n\to\mathbb R^m$ with clause coordinates $c_e$, the scalar function $\phi:\mathbb R\to\mathbb R_{\ge0}$, and the objective $\Phi:\mathbb R^n\to\mathbb R_{\ge0}$ by
\begin{equation}\label{eq:potential}
 c_e(z)\coloneqq z_u+z_v,\qquad
 \phi(c)\coloneqq \begin{cases}e^{-c}&c\ge0,\\1-c&c<0,\end{cases}
 \qquad \Phi(z)\coloneqq \sum_e\phi(c_e(z)).
\end{equation}
Both $\phi$ and $\Phi$ are convex and continuously differentiable. For each $z\in\mathbb R^n$, define the clause-weight vector $w(z)\in(0,1]^m$ coordinatewise by
\begin{equation}\label{eq:clauseweights}
 w_e(z)\coloneqq -\phi'(c_e(z))=\min\{1,e^{-c_e(z)}\}.
\end{equation}
This choice also gives $\phi(c)=\min\{1,e^{-c}\}+\pos{-c}$: besides the clause weight, $\Phi$ includes a term that will control violations caused by reassignment. We use Lemma~\ref{lem:randomthreshold}(i) with scores $-z_\ell$ to lower-bound $\tau_{w(z)}(F)$. Define the constant
\begin{equation}\label{eq:kdelta}
 k_\delta\coloneqq \frac{\delta}{4} e^{-4/\delta}.
\end{equation}
\begin{lemma}\label{lem:sublevel}
For every $z\in\mathbb R^n$ with $\Phi(z)\le m$,
\[
 L_{w(z)}(t)\ge2k_\delta m\quad\text{for every assignment }t\in\mathcal T,
 \qquad W(w(z))\ge2k_\delta m.
\]
\end{lemma}
\begin{proof}
We convert an assignment $t$ with small weighted loss $L_{w(z)}(t)$ into one that violates few clauses of $F$. The obstacle is that clauses with large positive $c_e(z)$ have very small weights, so their violations contribute little to $L_{w(z)}(t)$. Such a clause has an endpoint with large positive $z_\ell$; making that endpoint false satisfies the clause. We therefore reassign variables with large $|z_x|$, making their literal with positive $z_\ell$ false, and retain $t$ on the other variables. Clauses whose two variables retain $t$ then have weights bounded below. Randomizing the threshold lets us control violations involving reassignment using $\Phi(z)\le m$.

Fix $z\in\mathbb R^n$ with $\Phi(z)\le m$, an assignment $t\in\mathcal T$, and $r>0$. Choose $\theta$ uniformly in $[0,r]$ and define a random assignment $t^\theta\in\mathcal T$ as follows. On variables with $|z_x|\le\theta$, retain $t$. On every other variable $x$, make the literal $\ell\in\{x,\bar x\}$ with $z_\ell<0$ true and its complement false. This is consistent because $z_{\bar x}=-z_x$. This random assignment is used only in the proof; all probabilities and expectations below are over $\theta$.

For a clause $e=\{u,v\}$, we bound $\Pr_\theta[t^\theta_u=t^\theta_v=1]$, the probability that $t^\theta$ violates $e$, by separating two disjoint cases. First consider the event that $e$ is violated and both endpoint variables retain $t$. If this event is possible, then $t_ut_v=1$ and $|z_u|,|z_v|\le r$, so $c_e(z)\le2r$ and $w_e(z)\ge e^{-2r}$. Hence the probability of this event is at most $e^{2r}w_e(z)t_ut_v$.

For violations involving at least one reassigned variable, Lemma~\ref{lem:randomthreshold}(i), with $q_\ell=-z_\ell$ and $[\alpha,\beta]=[0,r]$, bounds their probability by $\pos{-c_e(z)}/r$. Combining the two cases gives
\[
 \Pr_\theta[t^\theta_u=t^\theta_v=1]
 \le e^{2r}w_e(z)t_ut_v+\frac{\pos{-c_e(z)}}r.
\]

Every assignment $t^\theta$ violates at least $\tau(F)$ clauses. Taking the expected number of violated clauses and using $\sum_e\pos{-c_e(z)}\le\Phi(z)\le m$ therefore gives
\[
 \delta m\le\tau(F)
 \le\E_\theta\!\left[\sum_{e=\{u,v\}}t^\theta_ut^\theta_v\right]
 \le e^{2r}L_{w(z)}(t)+\frac{m}r.
\]
Taking $r\coloneqq 2/\delta$ yields $L_{w(z)}(t)\ge(\delta/2)e^{-4/\delta}m=2k_\delta m$. The bound $W(w(z))\ge2k_\delta m$ follows from $L_{w(z)}(t)\le W(w(z))$.
\end{proof}

\subsection{Local evaluation of gradient descent}
We now reduce the imbalance $b(w(z))$ while keeping $\Phi(z)\le m$, so Lemma~\ref{lem:sublevel} applies at every iterate. Averaging the resulting clause weights will preserve the lower bound on $L_{w(z)}(t)$ for every assignment $t$ and give a deterministic local oracle.
To determine a step size for gradient descent, let $B\in\{-1,0,1\}^{n\times m}$ be the signed incidence matrix: $B_{xe}\coloneqq 1$ if $x\in e$, $-1$ if $\bar x\in e$, and zero otherwise. Each column has two nonzero entries, and each row has at most $\Delta$, so $\|B\|_2^2\le2\Delta$ by $(s+t)^2\le2s^2+2t^2$ on each column. The gradient is
\[
 \nabla\Phi(z)=B\phi'(B^{\mathsf T}z)=-b(w(z)),
\]
where $\phi'$ acts coordinatewise. Since $\phi'$ is one-Lipschitz on $\mathbb R$, for all $z,z'\in\mathbb R^n$,
\[
 \|\nabla\Phi(z)-\nabla\Phi(z')\|_2
 \le\|B\|_2^2\|z-z'\|_2\le2\Delta\|z-z'\|_2.
\]

Accordingly, starting at $z^0\coloneqq 0\in\mathbb R^n$, define the iterates $z^j\in\mathbb R^n$ for integers $j\ge0$ by the parallel updates
\begin{equation}\label{eq:update}
 z^{j+1}\coloneqq z^j+\frac{b(w(z^j))}{2\Delta}.
\end{equation}
The $2\Delta$-Lipschitz bound on $\nabla\Phi$ yields
\begin{equation}\label{eq:descent}
 \Phi(z^{j+1})\le\Phi(z^j)-\frac{\|b(w(z^j))\|_2^2}{4\Delta}.
\end{equation}
Thus every iterate has $\Phi(z^j)\le m$, and Lemma~\ref{lem:sublevel} gives $L_{w(z^j)}(t)\ge2k_\delta m$ for every assignment $t$. For an integer $T\ge1$, summing~\eqref{eq:descent} over $T$ rounds and using $\Phi\ge0$ gives
\begin{equation}\label{eq:sum}
 \sum_{j=0}^{T-1}\|b(w(z^j))\|_2^2\le4\Delta m.
\end{equation}
Equation~\eqref{eq:sum} controls the squared imbalances summed over rounds. To obtain one weight vector, we average the weights from all rounds; this avoids selecting a good round by inspecting its imbalance across the whole formula. Averaging also preserves the lower bound on $L_w(t)$ for every fixed assignment $t$, since $L_w(t)$ is linear in $w$. Define $\bar w\in(0,1]^m$ by
\begin{equation}\label{eq:averagedweights}
 \bar w\coloneqq \frac{1}T\sum_{j=0}^{T-1}w(z^j).
\end{equation}
To bound the imbalance of $\bar w$, note that $b$ is linear, so $\|b(w)\|_1$ is convex in $w$. There are at most $2m$ nonisolated variables; hence Cauchy--Schwarz and~\eqref{eq:sum} give
\begin{equation}\label{eq:mean}
 \|b(\bar w)\|_1\le\frac{1}T\sum_{j=0}^{T-1}\|b(w(z^j))\|_1
 \le\sqrt{\frac{2m}{T}\sum_{j=0}^{T-1}\|b(w(z^j))\|_2^2}
 \le m\sqrt{8\Delta/T}.
\end{equation}

We next show that the averaged weights $\bar w$ can be evaluated locally. The variable graph has the variables of $F$ as vertices, with two variables adjacent if they share a clause.
\begin{theorem}[Locally computable weights]\label{thm:local}
For constants $\delta>0$, $\Delta\ge2$, and $0<\beta\le1$, there is a deterministic local oracle specifying one weight vector $w\in(0,1]^m$: all query answers, including repeated queries, agree with this vector. Under the standing assumption $\tau(F)\ge\delta m>0$,
\[
 W(w)\ge k_\delta m,\qquad
 \frac{\|b(w)\|_1}{W(w)}\le\beta,\qquad
 \min_{t\in\mathcal T}\frac{L_w(t)}{W(w)}\ge k_\delta/2.
\]
A clause weight can be evaluated from the radius-$(T+1)$ neighborhood of its endpoints in the variable graph, where
\[
 T\coloneqq \left\lceil\frac{288\Delta}{k_\delta^2\beta^2}\right\rceil.
\]
The direct implementation uses $O(\Delta^{T+2})$ incidence queries per weight, depending only on $\delta,\Delta,\beta$.
\end{theorem}
\begin{proof}
Set $w\coloneqq \bar w$ from~\eqref{eq:averagedweights}. Since $W$ and $L_w(t)$ for each assignment $t$ are linear in $w$, averaging the bounds in Lemma~\ref{lem:sublevel} gives $W(w)\ge2k_\delta m$ and $L_w(t)\ge2k_\delta m$ for every $t$. By~\eqref{eq:mean} and the choice of $T$,
\[
 \frac{\|b(w)\|_1}{W(w)}\le\frac{1}{k_\delta}\sqrt{\frac{2\Delta}{T}}
 \le\beta/12\le\beta.
\]
Finally, $W(w)\le m$ gives $\min_{t\in\mathcal T} L_w(t)/W(w)\ge2k_\delta\ge k_\delta/2$.

It remains to evaluate $w$ locally. To determine the update at a variable in round $j+1$, query all its incident clauses and recursively compute the coordinates $z^j_\ell$ at their endpoints. For every $j<T$, these dependencies lie in the radius-$(T+1)$ neighborhood of the queried clause endpoints. Query this neighborhood once, compute the required coordinates in successive rounds, and accumulate $w_e(z^j)$ for $j=0,\ldots,T-1$.
The evaluation uses $O(\Delta^{T+2})$ queries. All answers agree with~\eqref{eq:averagedweights}, even when overlapping neighborhoods are evaluated separately. This evaluation requires neither the clause count $m$ nor preprocessing of the input.
\end{proof}
The tester will choose $\beta$ independently of $n$.

\section{Balanced Clause Distributions}\label{sec:comparison}
This section provides the exactly balanced clause distribution needed for the walk analysis in Section~\ref{sec:geo}. Normalize the weights from Theorem~\ref{thm:local} to $\pi\coloneqq w/W(w)$. We construct a balanced distribution $\widehat{\pi}$ for which $\tau_{\widehat{\pi}}(F)$ remains bounded below by a positive constant, while controlling the change from $\mu_\pi$ to $\mu_{\widehat{\pi}}$ in seed distributions and the change in conditional clause distributions (Theorem~\ref{thm:comparison}). These bounds will let us transfer the walk guarantees from $\widehat{\pi}$ back to $\pi$ in Section~\ref{sec:transfer}; the tester itself never computes $\widehat{\pi}$.

We use the induced literal distributions $\mu_\sigma$ and conditional clause distributions $q_\sigma(\cdot\mid\ell)$ from Section~\ref{sec:weighted-notation}.
We measure changes to the conditional clause distributions by their relative entropy $\KL$ (Section~\ref{sec:probability-notation}). We first define a convex penalty that combines these divergences, then optimize over balanced weights while penalizing such changes. The final subsection proves the rounding estimate in Lemma~\ref{lem:comparisonthreshold}, which we use to lower-bound the minimum weighted loss $\tau_{\widehat{\pi}}(F)$.

\subsection{Relative entropy of conditional clause distributions}
We now define the penalty for changing conditional clause distributions. It will be a weighted sum of their relative entropies; normalizing it will give the average divergence used to compare walks. Throughout this section, $F$ has $m$ E2SAT clauses, and $\pi\in\mathbb R_{>0}^m$ is a probability distribution on them. Write $\E_\pi$ for expectation over a clause $e=\{u,v\}\sim\pi$. We use the weighted degrees $d_w$, imbalance $b(w)$, and total imbalance $\|b(w)\|_1$ from~\eqref{eq:imbalance}, together with the induced literal distributions $\mu_\sigma$.

For a weight vector $v\in\mathbb R_{\ge0}^m$ and a literal $\ell$ with $d_v(\ell)>0$, we compare the distributions on clauses incident to $\ell$ that assign probabilities $v_e/d_v(\ell)$ and $\pi_e/(2\mu_\pi(\ell))$ to each such clause $e$, respectively. The walk comparison will use the average of these divergences over the literal distribution induced by $v$. Before normalizing $v$, the corresponding literal masses are $d_v(\ell)/2$. We therefore define $F_\pi:\mathbb R_{\ge0}^m\to\mathbb R_{\ge0}$ by
\begin{equation}\label{eq:conditional}
 F_\pi(v)\coloneqq\frac{1}{2}\sum_{\ell:d_v(\ell)>0}d_v(\ell)
 \KL\left(\left(\frac{v_e}{d_v(\ell)}\right)_{e\ni\ell}
 \middle\|\left(\frac{\pi_e}{2\mu_\pi(\ell)}\right)_{e\ni\ell}\right)\ge0.
\end{equation}
The coefficients $d_v(\ell)/2$ in~\eqref{eq:conditional} sum to $W(v)$. Thus, for $W(v)>0$, the average divergence over a literal drawn from $\mu_{v/W(v)}$ is $F_\pi(v)/W(v)$.

Expanding the conditional relative entropies gives
\begin{equation}\label{eq:F}
 F_\pi(v)=\sum_e v_e\log\frac{v_e}{\pi_e}
 -\frac{1}{2}\sum_\ell d_v(\ell)\log\frac{d_v(\ell)}{2\mu_\pi(\ell)}.
\end{equation}
The subtracted term removes the contribution from changing the literal masses, leaving only the weighted sum of relative entropies in~\eqref{eq:conditional}. Zero terms are interpreted by continuity. Any $\mu_\pi(\ell)=0$ also has $d_v(\ell)=0$ because $v$ is supported on the same clauses.

To verify convexity of $F_\pi$, consider a literal with $\mu_\pi(\ell)>0$. Its summand in~\eqref{eq:conditional} is one half of $\sum_{e\ni\ell}v_e\log\bigl(v_e/(d_v(\ell)q_\pi(e\mid\ell))\bigr)$. Both the numerator and the denominator inside each logarithm depend linearly on $v$, so the log-sum inequality gives convexity under mixtures of weight vectors. Summing over literals proves that $F_\pi$ is convex. Formula~\eqref{eq:F}, with the zero-term convention, also gives continuity on the nonnegative orthant and $F_\pi(cv)=cF_\pi(v)$ for $c\ge0$.

For a clause probability distribution $\widehat{\pi}\in[0,1]^m$, formula~\eqref{eq:F} becomes
\begin{equation}\label{eq:K1}
 F_\pi(\widehat{\pi})
 =\KL(\widehat{\pi}\|\pi)-\KL(\mu_{\widehat{\pi}}\|\mu_\pi).
\end{equation}
By~\eqref{eq:conditional},
\[
 F_\pi(\widehat{\pi})
 =\E_{\ell\sim\mu_{\widehat{\pi}}}
 \KL\bigl(q_{\widehat{\pi}}(\cdot\mid\ell)\|q_\pi(\cdot\mid\ell)\bigr).
\]
We will use $F_\pi(\widehat{\pi})$ to bound the divergence between walks in Lemma~\ref{lem:walkkl}.

\subsection{Constructing a balanced clause distribution}
We seek balanced weights $v\in\mathbb R_{\ge0}^m$ for which $\tau_v(F)$ remains bounded below by a positive constant while the conditional clause distributions change only slightly. Thus we want $\tau_v(F)$ to be large and $F_\pi(v)/W(v)$ to be small. We restrict to $0\le v\le\pi$, reducing the original weights to obtain balance. Once $W(v)$ is bounded below, this restriction will also control how much normalization can increase any clause probability.

For $\alpha>0$, we combine the two requirements in the objective $\alpha\tau_v(F)-F_\pi(v)$. A positive value implies
\[
 \frac{F_\pi(v)}{W(v)}<\alpha\frac{\tau_v(F)}{W(v)}\le\alpha,
\]
so smaller $\alpha$ requires smaller average divergence. More quantitatively, for $h>0$, a value at least $\alpha h$ implies $\tau_v(F)\ge h$ and $W(v)\ge h$. Since $W(v)\le1$, normalizing to $\widehat{\pi}=v/W(v)$ then gives $\tau_{\widehat{\pi}}(F)\ge h$ and $\widehat{\pi}_e\le\pi_e/h$. We therefore define the optimal value $V_\alpha(\pi)\in\mathbb R_{\ge0}$ by
\begin{equation}\label{eq:V}
 V_\alpha(\pi)\coloneqq \max_{0\le v\le\pi,\ b(v)=0}
 \{\alpha\tau_v(F)-F_\pi(v)\}.
\end{equation}
The maximum in~\eqref{eq:V} is attained by continuity and compactness, and $v=0$ gives $V_\alpha(\pi)\ge0$. We will prove a quantitative positive lower bound and normalize an optimizer. The theorem below gives the resulting guarantees with explicit constants.

\begin{theorem}[Balanced clause distribution]\label{thm:comparison}
For $0<\alpha\le1$ and $\beta\coloneqq \|b(\pi)\|_1$,
\begin{equation}\label{eq:roundbound}
 V_\alpha(\pi)\ge\frac{\alpha\,\tau_\pi(F)^2}{8}-\frac{\beta}{4}.
\end{equation}
In particular, if $\tau_\pi(F)\ge\delta>0$ and $\beta\le\alpha\delta^2/400$, then there is a balanced clause distribution $\widehat{\pi}\in[0,1]^m$ such that, with $h\coloneqq \delta^2/1600$,
\begin{equation}\label{eq:comparisonrho}
 \widehat{\pi}_e\le\frac{\pi_e}{2h},\quad
 \tau_{\widehat{\pi}}(F)\ge h,\quad
 F_\pi(\widehat{\pi})\le\alpha/2,\quad
 \KL(\mu_{\widehat{\pi}}\|\mu_\pi)\le\log\frac{1}{2h}.
\end{equation}
\end{theorem}
\begin{proof}
To lower-bound $V_\alpha(\pi)$ directly, we would need a single balanced weight vector whose loss is large for every assignment. We use minimax to let the weight vector respond to a fixed distribution of assignments, enforcing the balance constraints using Lagrange multipliers $z\in\mathbb R^n$. Extend them by $z_{\bar x}\coloneqq -z_x$, and recall $c_e(z)=z_u+z_v$. Let $\mathcal D:\mathcal T\to[0,1]$ range over all probability mass functions on $\mathcal T$. For a given $v$, the infimum over $z$ of $\sum_xz_xb_x(v)=\sum_ec_e(z)v_e$ is zero when $v$ is balanced and $-\infty$ otherwise. Also, $\inf_{\mathcal D}\E_{t\sim\mathcal D}L_v(t)=\tau_v(F)$. The box $0\le v\le\pi$ is compact and convex, and the objective in~\eqref{eq:comparisonminimax} is continuous, concave in $v$, and affine in $(\mathcal D,z)$. Minimax therefore gives
\begin{equation}\label{eq:comparisonminimax}
 V_\alpha(\pi)=\inf_{\mathcal D,z}\max_{0\le v\le\pi}
 \left\{\alpha\E_{t\sim\mathcal D}L_v(t)
       +\sum_ec_e(z)v_e-F_\pi(v)\right\}.
\end{equation}

It now suffices, for each fixed $\mathcal D,z$, to choose a trial weight vector $\widehat v$ in the box $0\le\widehat v\le\pi$ and lower-bound the objective in~\eqref{eq:comparisonminimax}. The trial vector need not be balanced: the outer infimum over $z$ already accounts for the balance constraint. Fix $\mathcal D,z$ and let $p\in[0,1]^m$ have coordinates $p_e\coloneqq \Pr_{t\sim\mathcal D}[t_u=t_v=1]$.

To construct $\widehat v$, we turn each multiplier $z_\ell$ into a bias $\xi_\ell\in(-1,1)$ by requiring $\log((1+\xi_\ell)/(1-\xi_\ell))=2z_\ell$, or equivalently $\xi_\ell=\tanh z_\ell$. The probabilities $(1+\xi_\ell)/2$ and $(1+\xi_{\bar\ell})/2$ sum to one. Consider a random assignment in which the variables are independent and each literal $\ell$ is true with probability $(1+\xi_\ell)/2$. Under this assignment, clause $\{u,v\}$ is violated with probability $(1+\xi_u)(1+\xi_v)/4$. We use this probability as the fraction of $\pi_e$ retained in $\widehat v_e$. The product form makes $\log(\widehat v_e/\pi_e)$ split into one term for $u$ and one for $v$, and the log-odds relation above lets us combine them with $c_e(z)=z_u+z_v$. Define $G\in(0,4)^m$ and $\widehat v\in\mathbb R_{\ge0}^m$ by
\begin{equation}\label{eq:trialweights}
 \xi_\ell=\tanh z_\ell,\qquad
 G_e\coloneqq (1+\xi_u)(1+\xi_v),\qquad
 \widehat v_e\coloneqq \frac{\pi_eG_e}{4}.
\end{equation}
Here $\xi_{\bar\ell}=-\xi_\ell$, and $0<\widehat v_e<\pi_e$.

To bound the part $\sum_ec_e(z)\widehat v_e-F_\pi(\widehat v)$ of the objective in~\eqref{eq:comparisonminimax}, apply the following identity for each literal $\ell$:
\begin{equation}\label{eq:trial-log-identity}
 z_\ell-\log\frac{1+\xi_\ell}{2}
 =-\frac{1}{2}\log\frac{1-\xi_\ell^2}{4}.
\end{equation}
In the resulting sum over literals, the denominator $2\mu_\pi(\ell)$ from~\eqref{eq:F} acquires the factor $(1-\xi_\ell^2)/4$. Define auxiliary literal masses $q\in\mathbb R_{\ge0}^{\mathcal L}$ by
\[
 q_\ell\coloneqq \frac{\mu_\pi(\ell)(1-\xi_\ell^2)}{2}.
\]
For every literal with $\mu_\pi(\ell)>0$, both $q_\ell$ and $d_{\widehat v}(\ell)$ are positive. Expanding~\eqref{eq:F} using~\eqref{eq:trial-log-identity} and the definition of $q_\ell$ gives
\begin{align}
 \sum_ec_e(z)\widehat v_e-F_\pi(\widehat v)
 &=\frac{1}{2}\sum_{\ell:\mu_\pi(\ell)>0}d_{\widehat v}(\ell)
       \log\frac{d_{\widehat v}(\ell)}{q_\ell}\notag\\
 &\ge W(\widehat v)-\frac{1}{2}\sum_\ell q_\ell\notag\\
 &=\frac{1}{8}\E_\pi(\xi_u+\xi_v)^2+\frac{1}{4}\sum_xb_x(\pi)\xi_x.\label{eq:trialentropy}
\end{align}
The inequality in~\eqref{eq:trialentropy} uses $x\log(x/y)\ge x-y$ for $x,y>0$. The last equality in~\eqref{eq:trialentropy} follows by expanding $G_e$ and using $\E_\pi(\xi_u+\xi_v)=\sum_xb_x(\pi)\xi_x$.

It remains to combine $\tfrac18\E_\pi(\xi_u+\xi_v)^2$ with the expected weighted loss $\alpha\E_{t\sim\mathcal D}L_{\widehat v}(t)$. Define $D_\xi,U\ge0$ by
\[
 D_\xi\coloneqq \E_\pi(\xi_u+\xi_v)^2,\qquad U\coloneqq \E_\pi[p_eG_e],
\]
so that the loss term is $\alpha\sum_ep_e\widehat v_e=\alpha U/4$. Lemma~\ref{lem:comparisonthreshold}, proved in the next subsection, gives $D_\xi+2U\ge\tau_\pi(F)^2$ for every assignment distribution $\mathcal D$ and every family $(\xi_\ell)_{\ell\in\mathcal L}$ with $\xi_\ell\in[-1,1]$ and $\xi_{\bar\ell}=-\xi_\ell$. Since $|\xi_x|\le1$, we also have $\sum_xb_x(\pi)\xi_x\ge-\beta$. Adding $\alpha U/4$ to~\eqref{eq:trialentropy} therefore yields
\begin{equation}\label{eq:trial-objective-bound}
 \alpha\sum_ep_e\widehat v_e+\sum_ec_e(z)\widehat v_e-F_\pi(\widehat v)
 \ge\frac{D_\xi+2\alpha U}{8}-\frac{\beta}{4}
 \ge\frac{\alpha\,\tau_\pi(F)^2}{8}-\frac{\beta}{4}.
\end{equation}
The last inequality in~\eqref{eq:trial-objective-bound} uses $\alpha\le1$ and $D_\xi+2U\ge\tau_\pi(F)^2$. The lower bound~\eqref{eq:trial-objective-bound} holds for every $\mathcal D,z$, so~\eqref{eq:comparisonminimax} proves~\eqref{eq:roundbound}.

To obtain the distribution guarantees in~\eqref{eq:comparisonrho}, suppose $\tau_\pi(F)\ge\delta$ and $\beta\le\alpha\delta^2/400$. Then~\eqref{eq:roundbound} gives $V_\alpha(\pi)\ge\alpha h$ for $h=\delta^2/1600$. Take an optimizer $v$ in~\eqref{eq:V} and put $W\coloneqq W(v)$. A uniformly random assignment violates each E2SAT clause with probability $1/4$, so $\tau_v(F)\le W/4\le W/2$. Together with $F_\pi(v)\ge0$ and $v\le\pi$, the inequality $\alpha\tau_v(F)-F_\pi(v)\ge\alpha h$ implies $2h\le W\le1$. Normalize $\widehat{\pi}\coloneqq v/W$. Then $\widehat{\pi}$ is balanced, and homogeneity gives
\[
 \tau_{\widehat{\pi}}(F)\ge h/W\ge h,\qquad
 F_\pi(\widehat{\pi})=\frac{F_\pi(v)}W
 \le\frac{\alpha\tau_v(F)}W\le\alpha/2.
\]
Finally, $v\le\pi$ and $W\ge2h$ give both $\widehat{\pi}_e\le\pi_e/(2h)$ and $\mu_{\widehat{\pi}}(\ell)\le\mu_\pi(\ell)/(2h)$. The bound on $\mu_{\widehat{\pi}}$ implies $\KL(\mu_{\widehat{\pi}}\|\mu_\pi)\le\log(1/(2h))$, completing~\eqref{eq:comparisonrho}.
\end{proof}
The distribution $\widehat{\pi}$ in~\eqref{eq:comparisonrho} is used only in the analysis. The tester samples with $w$ from Theorem~\ref{thm:local}.

\subsection{Reassigning variables using a random threshold}
We now prove the inequality $D_\xi+2U\ge\tau_\pi(F)^2$ used in the proof of Theorem~\ref{thm:comparison}. For the trial weights $\widehat v$ in~\eqref{eq:trialweights}, $U/4=\E_{t\sim\mathcal D}L_{\widehat v}(t)$ is the expected weighted loss under $\widehat v$, and $D_\xi/8$ is the first term in the final expression of~\eqref{eq:trialentropy}. The next lemma applies to any $\xi_\ell\in[-1,1]$ satisfying $\xi_{\bar\ell}=-\xi_\ell$, including $\xi_\ell=\tanh z_\ell$. In the proof, we reassign variables using a random threshold and bound the expected weighted loss $\E L_\pi(t^\theta)$ of the resulting assignment $t^\theta$ by $U+\sqrt{D_\xi}$.

\begin{lemma}[Reassignment using a random threshold]\label{lem:comparisonthreshold}
Let $\mathcal D:\mathcal T\to[0,1]$ be a probability mass function on $\mathcal T$, and let $p\in[0,1]^m$ have coordinates $p_e\coloneqq \Pr_{t\sim\mathcal D}[t_u=t_v=1]$. For each literal $\ell$, let $\xi_\ell\in[-1,1]$, with $\xi_{\bar\ell}=-\xi_\ell$. Define a vector $G\in[0,4]^m$ and nonnegative scalars $D_\xi,U$ by
\[
 G_e\coloneqq (1+\xi_u)(1+\xi_v),\qquad
 D_\xi\coloneqq \E_\pi(\xi_u+\xi_v)^2,\qquad U\coloneqq \E_\pi[p_eG_e].
\]
Here $U$ is the expected loss of $t\sim\mathcal D$ under clause weights $\pi_eG_e$, while $D_\xi$ is the mean square of $\xi_u+\xi_v$ for $\{u,v\}\sim\pi$. Then
\begin{equation}\label{eq:thresholdloss}
 \tau_\pi(F)\le U+\sqrt{D_\xi},\qquad
 \tau_\pi(F)^2\le D_\xi+2U.
\end{equation}
\end{lemma}
\begin{proof}
Draw $t\sim\mathcal D$ and, independently, $\theta$ uniformly from $[0,1]$. On variables with $|\xi_x|\le\theta$, keep $t$; on the remaining variables, make the literal $\ell$ with $\xi_\ell>0$ true. Since $\xi_{\bar\ell}=-\xi_\ell$, this defines an assignment $t^\theta\in\mathcal T$.

For a clause $e=\{u,v\}$, put $M_e\coloneqq \max\{|\xi_u|,|\xi_v|\}\in[0,1]$. The probability that $e$ is violated and both endpoint variables retain $t$ is $p_e(1-M_e)$. Lemma~\ref{lem:randomthreshold}(i), with $q_\ell=\xi_\ell$ and $[\alpha,\beta]=[0,1]$, bounds the probability of a violation involving reassignment by $\pos{\xi_u+\xi_v}$.

Since $|\xi_u\xi_v|\le M_e$, we have
\[
 1-M_e\le1+\xi_u\xi_v
 =G_e-(\xi_u+\xi_v)\le G_e+\pos{-\xi_u-\xi_v}.
\]
Using $p_e\le1$ therefore gives
\[
 \Pr[t^\theta_u=t^\theta_v=1]
 \le p_e(1-M_e)+\pos{\xi_u+\xi_v}
 \le p_eG_e+|\xi_u+\xi_v|.
\]
Summing against $\pi$ and applying Cauchy--Schwarz yields
$\tau_\pi(F)\le\E L_\pi(t^\theta)\le U+\sqrt{D_\xi}$.
For the second claim, put $\tau\coloneqq \tau_\pi(F)\le1$. Then
$D_\xi\ge\pos{\tau-U}^2\ge\tau^2-2U$, proving~\eqref{eq:thresholdloss}.
\end{proof}

\section{Walks from Complementary Literals}\label{sec:geo}
We prove a constant lower bound on the two-way affinity $B_{\widehat{\pi}}$ introduced in Section~\ref{sec:overview}, for walks of logarithmic expected length (Theorem~\ref{thm:geo}). The bound applies when $\widehat{\pi}$ is balanced and $\tau_{\widehat{\pi}}(F)$ is bounded below by a positive constant. This supplies the guarantee that Section~\ref{sec:tester} will transfer to the tester's walks and turn into contradictory implication paths.
The main difficulty is to obtain the implication paths $s\leadsto\bar s$ and $\bar s\leadsto s$ for the same literal $s$. Our test obtains the first path from complementary endpoint collisions among walks started at $s$, and the reverse path from such collisions among walks started at $\bar s$. We therefore analyze the two tests through $B_{\widehat{\pi}}$, which averages the minimum of the endpoint affinities from $s$ and $\bar s$. A lower bound on the affinity averaged over all starting literals does not give a lower bound on $B_{\widehat{\pi}}$.
We therefore seek a set where complementary literals have close affinities, so replacing each affinity by their minimum loses little. It then suffices to lower-bound this set's contribution to the average affinity. We obtain that lower bound by applying a general lemma to the trace chain on the set and transferring the bound to the original walks.

\paragraph{Walk notation.}
We first define the walks and affinities used here and in Section~\ref{sec:tester}, allowing any clause probability distribution $\sigma\in[0,1]^m$. We use $\mu_\sigma$ and $q_\sigma$ from Section~\ref{sec:weighted-notation}. On any complement-closed set $\Omega\subseteq\mathcal L$, let $I$ be the identity on $\mathbb R^\Omega$ and let $C:\mathbb R^\Omega\to\mathbb R^\Omega$ be the \emph{complementation operator}, $(Cf)(\ell)\coloneqq f(\bar\ell)$. If $p:\Omega\to[0,1]$ is a probability mass function, then $Cp$ is the law of the complemented literal.

Let $T_\sigma\in[0,1]^{\mathcal L\times\mathcal L}$ be the \emph{forward implication kernel}. From a literal $\ell$ with $\mu_\sigma(\ell)>0$, it chooses a clause $e=\{\ell,v\}$ according to $q_\sigma(\cdot\mid\ell)$ and follows the implication $\ell\to\bar v$. At a literal with $\mu_\sigma(\ell)=0$, it stays at $\ell$. The \emph{lazy implication kernel} is $P_\sigma\coloneqq(I+T_\sigma)/2\in[0,1]^{\mathcal L\times\mathcal L}$.

To treat both these walks and the trace chain below, we define \emph{geometric stopping} for any row-stochastic kernel $K\in[0,1]^{\Omega\times\Omega}$ on a complement-closed literal set $\Omega$. Independently stop with probability $\eta\in(0,1)$ before each step of $K$. The \emph{endpoint kernel} $R^K_\eta\in[0,1]^{\Omega\times\Omega}$ and its affinity function $a^K_\eta:\Omega\to[0,1]$ are
\begin{equation}\label{eq:geometrickernel}
 R^K_\eta\coloneqq\eta\sum_{r\ge0}(1-\eta)^rK^r,\qquad
 a^K_\eta(s)\coloneqq\Aff(R^K_\eta(s,\cdot),CR^K_\eta(s,\cdot)).
\end{equation}
The walk takes exactly $r$ steps with probability $\eta(1-\eta)^r$, and the stopping rule gives $R^K_\eta=\eta I+(1-\eta)KR^K_\eta$. We use this recursion to control entropy dissipation and the variation of affinities. The survival factor $(1-\eta)^r$ will also let us compare walks observed at successive returns to a subset.

For the lazy implication kernel $P_\sigma$, write $R_{\sigma,\eta}\in[0,1]^{\mathcal L\times\mathcal L}$ and $a_{\sigma,\eta}:\mathcal L\to[0,1]$ for its endpoint kernel and affinity function, and define the two-way affinity $B_{\sigma,\eta}\in[0,1]$ by
\begin{equation}\label{eq:distributionaffinity}
 \begin{aligned}
 R_{\sigma,\eta}&\coloneqq R^{P_\sigma}_\eta,\qquad
 a_{\sigma,\eta}\coloneqq a^{P_\sigma}_\eta,\\
 B_{\sigma,\eta}&\coloneqq\E_{s\sim\mu_\sigma}
       \min\{a_{\sigma,\eta}(s),a_{\sigma,\eta}(\bar s)\}.
 \end{aligned}
\end{equation}
The subscripts specify the clause distribution and stopping probability; the argument of $a_{\sigma,\eta}$ specifies the starting literal. When $\eta$ is understood, we omit it from the subscripts of $a_{\sigma,\eta}$ and $B_{\sigma,\eta}$.

The affinity $a_{\sigma,\eta}(s)$ measures the overlap between the endpoint distribution from $s$ and its complement. If two independent forward walks from $s$ end at $z$ and $\bar z$, reversing and complementing the second path gives $z\leadsto\bar s$. Concatenating it with the first path gives $s\leadsto\bar s$. The walks have independent geometric lengths. A contradiction also requires the path $\bar s\leadsto s$, obtained by the same test from $\bar s$, so $B_{\sigma,\eta}$ takes the minimum of the two affinities before averaging.

\paragraph{Specialization to a balanced clause distribution.}
For the remainder of this section, fix a balanced clause distribution $\widehat{\pi}$ and write $\mu\coloneqq\mu_{\widehat{\pi}}$. We work on its positive support $\mathcal L_+\coloneqq\{\ell\in\mathcal L:\mu(\ell)>0\}$. Balance makes $\mathcal L_+$ closed under complementation and under the transitions of $T_{\widehat{\pi}}$. We may therefore restrict $\mu$ and the kernels $T_{\widehat{\pi}},P_{\widehat{\pi}},R_{\widehat{\pi},\eta}$ to $\mathcal L_+$, abbreviating the kernels to $T,P,R_\eta\in[0,1]^{\mathcal L_+\times\mathcal L_+}$, respectively. Define $a:\mathcal L_+\to[0,1]$ by $a(s)\coloneqq a_{\widehat{\pi},\eta}(s)$. Since $\mu$ is supported on $\mathcal L_+$, this restriction leaves $B_{\widehat{\pi}}$ unchanged. Balance gives $\mu T=\mu$, so $\mu$ is stationary for $T$ and $P$. Set $N\coloneqq2n$, so that $|\mathcal L_+|\le N$.

\begin{theorem}\label{thm:geo}
For every $0<\delta\le1$, there are constants $b_\delta,C_\delta>0$ and a choice of stopping probability $\eta=\eta(\delta,n)\in(0,1)$ with
\[
 \eta^{-1}\le C_\delta\log(2n)
\]
such that every clause distribution $\widehat{\pi}$ with $b(\widehat{\pi})=0$ and $\tau_{\widehat{\pi}}(F)\ge\delta$ satisfies $B_{\widehat{\pi},\eta}\ge b_\delta$. The constants and the choice of $\eta$ are explicit in the proof below.
\end{theorem}
\paragraph{Proof strategy.}
A lower bound on $\E_{s\sim\mu}a(s)$ alone does not control $B_{\widehat{\pi}}$: numerical pairs $(a(s),a(\bar s))=(1,0)$ would have positive average affinity but zero minimum. We will use Lemma~\ref{lem:oneway} to lower-bound the average affinity of the trace chain on a set where $a(s)$ and $a(\bar s)$ are close. Lemma~\ref{lem:comparison} transfers this estimate to a lower bound on that set's contribution to the original average.

Specifically, we will choose a constant $0<\lambda<1/2$ depending only on $\delta$ and a complement-closed set $S\subseteq\mathcal L_+$ on which $|a(s)-a(\bar s)|\le2\lambda$. Our goal is to obtain
\begin{equation}\label{eq:geostrategy}
 4\lambda\ \le\ \sum_{s\in S}\mu(s)a(s)\ \le\ 2\lambda+B_{\widehat{\pi}},
\end{equation}
which gives $B_{\widehat{\pi}}\ge2\lambda$. The upper bound follows from $a(s)\le2\lambda+\min\{a(s),a(\bar s)\}$ on $S$ and $\mu(S)\le1$. For the lower bound, we will retain positive minimum weighted loss within $S$, measured by $\min_{t\in\mathcal T}\sum_{e=\{u,v\}\subseteq S}\widehat{\pi}_e t_ut_v$ in the original weights $\widehat{\pi}$.

To construct $S$, Lemma~\ref{lem:stability} bounds the mean squared change of $a(\bar s)-a(s)$ along implications. We assign variables $x$ for which $a(x)$ and $a(\bar x)$ differ by more than a threshold, making the literal $\ell$ with $a(\bar\ell)>a(\ell)$ true. A violated clause all of whose variables are assigned forces large positive differences at both endpoints. Averaging over the threshold controls the weight of clauses with exactly one endpoint in $S$, the set of literals of unassigned variables. Lemma~\ref{lem:threshold} gives a threshold for which the total $\widehat{\pi}$-weight of these clauses and violated clauses with both endpoints outside $S$ is at most $\delta/2$. Thus every assignment on $S$ still violates clauses with both endpoints in $S$ of total $\widehat{\pi}$-weight at least $\delta/2$, and $\mu(S)\ge\delta/2$.

To apply Lemma~\ref{lem:oneway} on $S$, we observe $P$ at successive returns to $S$. The trace chain retains the original transitions within $S$ and has stationary distribution $\mu$ conditioned on $S$, which is equal at complementary literals (Lemma~\ref{lem:trace}). These properties let us apply the average-affinity bound using the original internal clauses. When the starting state is drawn from $\mu$ conditioned on $S$, Lemma~\ref{lem:trace} also bounds the expected return time by $1/\mu(S)$. In Lemma~\ref{lem:comparison}, we use this bound to control the mass lost to stopping along the expanded paths. By choosing a smaller stopping probability for $P$ than for the trace chain, we transfer the affinity bound back to $P$, obtaining the lower bound in~\eqref{eq:geostrategy}. The proof of Theorem~\ref{thm:geo} chooses $\lambda$ and both stopping probabilities using only $\delta,n$, before constructing $S$.

\paragraph{Walk identities used in the proof.}
As an operator on functions, $T:\mathbb R^{\mathcal L_+}\to\mathbb R^{\mathcal L_+}$ acts by $(Tf)(i)=\sum_jT(i,j)f(j)$; probability distributions are represented by row vectors in expressions such as $\mu T$. The \emph{stationary flow} of a labeled arc is the probability of traversing it when the starting literal has distribution $\mu$. Each of the two labeled arcs $u\to\bar v$ and $v\to\bar u$ has stationary flow $\widehat{\pi}_e/2$. Let $T^*,P^*\in[0,1]^{\mathcal L_+\times\mathcal L_+}$ be the \emph{reverse kernels} of $T,P$ with respect to $\mu$, so $T^*(i,j)=\mu(j)T(j,i)/\mu(i)$. The complementary arc pairs give
\[
 T^*=CTC,\qquad P^*=(I+T^*)/2=CPC.
\]
Consequently, $CR_\eta(s,\cdot)$ is also the endpoint distribution of a reverse walk from $\bar s$. This identity is used only in the analysis; the collision test samples forward walks as described above.

\subsection{Average endpoint affinity}\label{sec:average-affinity}
We lower-bound the average endpoint affinity by rounding vectors constructed from the endpoint distributions, following the entropy dissipation and rounding approach of Yoshida~\cite{manuscript}. To use the same argument for the trace chain, the next lemma allows clause weights that need not be balanced or sum to one. It requires a lazy chain with a stationary distribution equal at complementary literals and lower bounds on its transitions along implications. These conditions will let us use the original internal clauses on $S$.

\begin{lemma}\label{lem:oneway}
Let $\Omega\subseteq\mathcal L$ be complement closed, with $|\Omega|\le N$, and let $\nu:\Omega\to(0,1]$ be a probability mass function satisfying $\nu(\ell)=\nu(\bar\ell)$. Let $K\in[0,1]^{\Omega\times\Omega}$ be a row-stochastic kernel with stationary distribution $\nu$ and $K(\ell,\ell)\ge1/2$. Then $J\coloneqq 2K-I\in[0,1]^{\Omega\times\Omega}$ is also a row-stochastic kernel with stationary distribution $\nu$. Let $\omega\in\mathbb R_{\ge0}^m$ be clause weights, zero unless both endpoints lie in $\Omega$, such that
\begin{equation}\label{eq:flowdomination}
 \sum_{e=\{i,\bar j\}}\omega_e\le2\nu(i)J(i,j)
 \qquad(i,j\in\Omega).
\end{equation}
For $0<\kappa<1$, define the average endpoint affinity $A\in[0,1]$ by
\[
 A\coloneqq\E_{s\sim\nu}a^K_\kappa(s),
\]
using~\eqref{eq:geometrickernel} with stopping probability $\kappa$. If $\tau_\omega(F)\ge\delta>0$, then
\[
 A\ge\delta^2-2\kappa\log_2 N.
\]
\end{lemma}
\begin{proof}
Summing~\eqref{eq:flowdomination} over $j$ gives $d_\omega(i)\le2\nu(i)$, and hence $W(\omega)\le1$. We will construct an assignment with weighted loss at most $\tfrac12\sqrt{\kappa\log_2N+A}$.

To this end, for each seed $s\in\Omega$ let $f_s,f_s^0:\Omega\to\mathbb R_{\ge0}$ be the densities of $R^K_\kappa(s,\cdot)$ and the point mass at $s$, respectively, with respect to $\nu$:
\[
 f_s(z)\coloneqq \frac{R^K_\kappa(s,z)}{\nu(z)},\qquad
 f_s^0(z)\coloneqq \frac{\mathbf 1_{\{z=s\}}}{\nu(s)}.
\]

First, we bound the mean squared variation of the square roots of these densities along transitions of $J$. For a density $f:\Omega\to\mathbb R_{\ge0}$ with $\E_\nu f=1$, define
$\operatorname{Ent}_\nu(f)\coloneqq \E_\nu f\log_2 f$, with $0\log_2 0=0$, and
$\mathcal E_J(f)\coloneqq \tfrac12\E_{X\sim\nu,Y\sim J(X,\cdot)}(\sqrt{f(X)}-\sqrt{f(Y)})^2$.
Let $J^*,K^*\in[0,1]^{\Omega\times\Omega}$ be the reverse kernels of $J,K$ with respect to $\nu$; then $K^*=(I+J^*)/2$.
For $a,b\ge0$, the scalar inequality
\[
 a\log_2a+b\log_2b-(a+b)\log_2((a+b)/2)
 \ge a+b-2\sqrt{ab}
\]
will be applied with $a=f(X)$ and $b=f(Y)$, where $X\sim\nu$ and, conditional on $X$, $Y\sim J^*(X,\cdot)$. Define $\psi:\mathbb R_{\ge0}\to\mathbb R$ by $\psi(t)\coloneqq t\log_2t$ for $t>0$ and $\psi(0)\coloneqq0$. Since $(K^*f)(X)=\E[(f(X)+f(Y))/2\mid X]$, conditional Jensen's inequality gives
\[
 \operatorname{Ent}_\nu(K^*f)
 =\E\psi\!\left(\E\!\left[\frac{f(X)+f(Y)}{2}\,\middle|\,X\right]\right)
 \le\E\psi\!\left(\frac{f(X)+f(Y)}{2}\right).
\]
Both $X$ and $Y$ have distribution $\nu$, so $\operatorname{Ent}_\nu(f)=\tfrac12\E[\psi(f(X))+\psi(f(Y))]$. Subtracting the preceding upper bound and then applying the scalar inequality yields
\begin{align*}
 \operatorname{Ent}_\nu(f)-\operatorname{Ent}_\nu(K^*f)
 &\ge\frac{1}{2}\E\!\left[\psi(f(X))+\psi(f(Y))
       -2\psi\!\left(\frac{f(X)+f(Y)}{2}\right)\right]\\
 &\ge\frac{1}{2}\E\bigl(\sqrt{f(X)}-\sqrt{f(Y)}\bigr)^2.
\end{align*}
Reversing the stationary flow exchanges $J^*$ with $J$ and preserves the squared difference. Thus
\begin{equation}\label{eq:entropydissipation}
 \mathcal E_J(f)\le\operatorname{Ent}_\nu(f)-\operatorname{Ent}_\nu(K^*f).
\end{equation}
For the stopped walk, the identity $R^K_\kappa=\kappa I+(1-\kappa)R^K_\kappa K$ gives $f_s=\kappa f_s^0+(1-\kappa)K^*f_s$. Convexity and nonnegativity of relative entropy imply
\[
 \operatorname{Ent}_\nu(f_s)-\operatorname{Ent}_\nu(K^*f_s)
 \le\kappa\bigl(\operatorname{Ent}_\nu(f_s^0)-\operatorname{Ent}_\nu(K^*f_s)\bigr)
 \le\kappa\log_2\frac{1}{\nu(s)}.
\]
Combining this with~\eqref{eq:entropydissipation} and averaging over $s\sim\nu$ gives
\begin{equation}\label{eq:energy}
 \E_{s\sim\nu}\mathcal E_J(f_s)
 \le\kappa\sum_s\nu(s)\log_2\frac{1}{\nu(s)}
 \le\kappa\log_2 N.
\end{equation}

To use this bound for rounding, define vectors $v_\ell\in\mathbb R^\Omega$ for $\ell\in\Omega$ by
\[
 v_\ell(s)\coloneqq \sqrt{\nu(s)f_s(\ell)}.
\]
Stationarity gives $\|v_\ell\|=1$. Let $c_\ell\coloneqq \ip{v_\ell}{v_{\bar\ell}}\in[0,1]$, so $c_\ell=c_{\bar\ell}$. The identity $\nu(\ell)=\nu(\bar\ell)$ gives $\E_\nu c_\ell=A$.

To transfer~\eqref{eq:energy} to these vectors, expand the squared norm in the seed coordinate $s$: for any $i,j\in\Omega$,
\[
 \|v_i-v_j\|^2
 =\sum_{s\in\Omega}\nu(s)
   \bigl(\sqrt{f_s(i)}-\sqrt{f_s(j)}\bigr)^2.
\]
Taking expectation over $X\sim\nu$, $Y\sim J(X,\cdot)$ and exchanging it with the sum over $s$ gives
\begin{equation}\label{eq:vectorvariation}
 \begin{aligned}
 \E_{X\sim\nu,\,Y\sim J(X,\cdot)}\|v_X-v_Y\|^2
 &=2\E_{s\sim\nu}\mathcal E_J(f_s)\\
 &\le2\kappa\log_2N.
 \end{aligned}
\end{equation}
Here the equality uses the factor $1/2$ in the definition of $\mathcal E_J$, and the inequality is~\eqref{eq:energy}.

For consistent Gaussian rounding, we need unit vectors that are negatives of each other at complementary literals. The differences $v_\ell-v_{\bar\ell}$ have the required symmetry but varying norms. Complete them to unit vectors by defining $X_x,X_{\bar x}\in\mathbb R^\Omega\oplus\mathbb R$ for each variable $x$ with $x\in\Omega$:
\[
 X_x\coloneqq \left(\frac{v_x-v_{\bar x}}{\sqrt2},\sqrt{c_x}\right),
 \qquad X_{\bar x}\coloneqq -X_x.
\]
Indeed, $\|X_x\|^2=(2-2c_x)/2+c_x=1$. For a clause $\{u,v\}$, the first coordinates of $X_u+X_v$ equal $((v_u-v_{\bar v})+(v_v-v_{\bar u}))/\sqrt2$, and the square of its last coordinate is at most $2(c_u+c_v)$. Hence
\[
 \|X_u+X_v\|^2
 \le\|v_u-v_{\bar v}\|^2+\|v_v-v_{\bar u}\|^2+2(c_u+c_v).
\]
Each clause supplies two implication arcs, so~\eqref{eq:flowdomination} and~\eqref{eq:vectorvariation} give
\[
 \sum_{e=\{u,v\}}\omega_e
 \bigl(\|v_u-v_{\bar v}\|^2+\|v_v-v_{\bar u}\|^2\bigr)
 \le2\E_{X\sim\nu,\,Y\sim J(X,\cdot)}\|v_X-v_Y\|^2
 \le4\kappa\log_2N.
\]
Also, $\sum_e\omega_e(c_u+c_v)=\sum_\ell d_\omega(\ell)c_\ell\le2\E_\nu c_\ell=2A$. Combining these bounds yields
\begin{equation}\label{eq:roundvectors}
 \sum_{e=\{u,v\}}\omega_e\|X_u+X_v\|^2\le4\kappa\log_2N+4A.
\end{equation}
To complete the rounding, draw a standard Gaussian vector in $\mathbb R^\Omega\oplus\mathbb R$, and make literal $\ell$ true when its inner product with $X_\ell$ is positive. Since $X_{\bar\ell}=-X_\ell$, this gives a consistent assignment with probability one. The direction of the Gaussian projection onto a plane containing $X_u$ and $X_v$ is uniform. Writing $\theta\coloneqq \angle(X_u,-X_v)\in[0,\pi]$, the directions for which both inner products are positive form a sector of angle $\theta$. Thus the probability of violating $\{u,v\}$ is $\theta/(2\pi)$. Since the vectors are unit vectors,
\[
 \|X_u+X_v\|=2\sin(\theta/2)\ge2\theta/\pi,
 \qquad \frac{\theta}{2\pi}\le\frac{\|X_u+X_v\|}{4}.
\]
By Cauchy--Schwarz,~\eqref{eq:roundvectors}, and $W(\omega)\le1$, some assignment $t\in\mathcal T$ therefore satisfies
\[
 L_\omega(t)\le\frac{1}{4}\sum_e\omega_e\|X_u+X_v\|
 \le\frac{1}{2}\sqrt{W(\omega)(\kappa\log_2N+A)}
 \le\frac{1}{2}\sqrt{\kappa\log_2N+A}.
\]
Since $\tau_\omega(F)\ge\delta$, we obtain
\[
 A\ge4\delta^2-\kappa\log_2N\ge\delta^2-2\kappa\log_2N.
\]
\end{proof}
Taking $\Omega=\mathcal L_+$, $K=P$, $\nu=\mu$, $\omega=\widehat{\pi}$, and $\kappa=\eta$ gives equality in~\eqref{eq:flowdomination}. Thus Lemma~\ref{lem:oneway} yields $\E_{s\sim\mu}a(s)\ge\delta^2-2\eta\log_2N$ whenever $\tau_{\widehat{\pi}}(F)\ge\delta$.

\subsection{Assigning variables with unequal affinities}
We now construct the set $S$ used in~\eqref{eq:geostrategy}. We assign variables where $a(x)$ and $a(\bar x)$ differ substantially, while bounding the total $\widehat{\pi}$-weight of violated clauses all of whose variables are assigned and clauses with one assigned and one unassigned variable. The clauses all of whose variables remain unassigned will then have positive minimum weighted loss. For each remaining variable $x$, the affinities $a(x)$ and $a(\bar x)$ will be nearly equal, yielding the upper bound in~\eqref{eq:geostrategy}.

For the affinity function $a$ in~\eqref{eq:distributionaffinity}, define the difference vector $d\in[-1,1]^{\mathcal L_+}$ by $d_s\coloneqq a(\bar s)-a(s)$. Since $d_{\bar s}=-d_s$, the change in $d$ along an implication $u\to\bar v$ is $d_u-d_{\bar v}=d_u+d_v$. We therefore measure its mean squared change by the nonnegative scalar $D$ defined as
\[
 D\coloneqq \E_{\{u,v\}\sim\widehat{\pi}}(d_u+d_v)^2.
\]
The next lemma bounds $D$ in terms of the stopping probability $\eta$. This will let us assign variables with large $|d_s|$ by making their literal with positive $d$ true: a violated clause all of whose variables are assigned has both $d_u$ and $d_v$ positive and large, so it contributes substantially to $D$.

\begin{lemma}\label{lem:stability}
For $d$ and $D$ defined above, if $\eta\le1/2$, then
\[
 d_{\bar s}=-d_s,\qquad
 D\le32\eta.
\]
\end{lemma}
\begin{proof}
Separating the zero-step term in~\eqref{eq:geometrickernel} gives $R_\eta=\eta I+(1-\eta)P R_\eta$. Concavity of affinity then implies
$a(s)\ge(1-\eta)(Pa)(s)$, since the point mass at $s$ has zero affinity with its complement. With $c\coloneqq \eta/(1-\eta)$, this gives $Pa-a\le ca\le c$ pointwise. For functions $f,g\in\mathbb R^{\mathcal L_+}$, define their $\mu$-weighted inner product by $\ip{f}{g}_\mu\coloneqq \sum_s\mu(s)f_sg_s$. Stationarity and $0\le a(s)\le1$ give
\[
 \ip{a}{(I-P)a}_\mu
 =\ip{1-a}{(P-I)a}_\mu\le c.
\]
Therefore $\E_{X\sim\mu,\,Y\sim T(X,\cdot)}(a(X)-a(Y))^2\le4c$.
Because $T^*=CTC$, replacing $a$ by $Ca$ in this mean square is equivalent to replacing $T$ by $T^*$. Swapping $X$ and $Y$ changes their joint distribution from $\mu(X)T(X,Y)$ to $\mu(X)T^*(X,Y)$ but leaves $(a(X)-a(Y))^2$ unchanged. Thus the same upper bound $4c$ applies to $Ca$. Applying $(x-y)^2\le2x^2+2y^2$ to $d=Ca-a$ gives
\[
 \E_{X\sim\mu,\,Y\sim T(X,\cdot)}(d_X-d_Y)^2
 \le16c\le32\eta.
\]
For each clause $\{u,v\}$, both implication arcs contribute $(d_u+d_v)^2$ because $d_{\bar\ell}=-d_\ell$. Thus the last expectation equals $D$.
\end{proof}

To retain positive minimum weighted loss among clauses all of whose variables are unassigned, we must also control clauses with one assigned and one unassigned variable. Lemma~\ref{lem:randomthreshold}(ii) bounds the probability of this event together with violations among assigned variables. We apply it with scores $d_\ell$ and average over clauses to bound the combined $\widehat{\pi}$-weight in terms of $D$.

\begin{lemma}[Partial assignment]\label{lem:threshold}
Assume $\tau_{\widehat{\pi}}(F)\ge\delta>0$. With $d_\ell=a(\bar\ell)-a(\ell)$ and $D=\E_{\widehat{\pi}}(d_u+d_v)^2$ as in Lemma~\ref{lem:stability}, for any $0<\lambda<1/2$ there is a threshold $\theta\in[\lambda,2\lambda]$ such that the complement-closed set
\[
 S\coloneqq \{\ell\in\mathcal L_+:|d_\ell|\le\theta\}
\]
has the following properties. Assign every variable $x$ with $x\in\mathcal L_+\setminus S$ by making its literal $\ell$ with $d_\ell>0$ true. The total $\widehat{\pi}$-weight of clauses with exactly one endpoint in $S$ and violated clauses with both endpoints outside $S$ is at most
\[
 e_0\coloneqq \frac{\sqrt D}{\lambda}.
\]
In the following minimum, $t$ ranges over vectors in $\{0,1\}^S$ satisfying $t_{\bar x}=1-t_x$ for each variable of $S$. If $e_0\le\delta/2$, then
\[
 \min_{t\text{ on the variables of }S}
 \sum_{e=\{u,v\}\subseteq S}\widehat{\pi}_e t_ut_v\ge\delta/2,
 \qquad \mu(S)\ge\delta/2.
\]
This sum uses the original weights $\widehat{\pi}$, without renormalizing the weights of clauses $\{u,v\}\subseteq S$. Moreover,
\begin{equation}\label{eq:smallS}
 \sum_{\ell\in S}\mu(\ell)a(\ell)\le2\lambda+B_{\widehat{\pi}}.
\end{equation}
\end{lemma}
\begin{proof}
Choose $\theta$ uniformly in $[\lambda,2\lambda]$. Apply Lemma~\ref{lem:randomthreshold}(ii) with $q_\ell=d_\ell$ on $\mathcal L_+$, $q_\ell=0$ elsewhere, and $[\alpha,\beta]=[\lambda,2\lambda]$. For each clause $\{u,v\}$ of positive $\widehat{\pi}$-weight, the probability that exactly one endpoint lies in $S$, or both lie outside $S$ and the clause is violated, is at most $|d_u+d_v|/\lambda$. Averaging over clauses and applying Cauchy--Schwarz gives a threshold for which the total $\widehat{\pi}$-weight of these clauses is at most $\E_{\widehat{\pi}}|d_u+d_v|/\lambda\le\sqrt D/\lambda=e_0$.

Choose such a threshold and its set $S$. Extending any assignment on $S$ by the prescribed outside assignment gives a full assignment $t\in\mathcal T$ with $L_{\widehat{\pi}}(t)\le\sum_{e=\{u,v\}\subseteq S}\widehat{\pi}_e t_ut_v+e_0$. If $e_0\le\delta/2$, the assumption $\tau_{\widehat{\pi}}(F)\ge\delta$ therefore forces this sum to be at least $\delta-e_0\ge\delta/2$ for every assignment on $S$. The total $\widehat{\pi}$-weight of clauses entirely in $S$ is at most $\mu(S)$, so $\mu(S)\ge\delta/2$ as well. Finally, for $\ell\in S$,
$a(\ell)\le|d_\ell|+\min(a(\ell),a(\bar\ell))\le2\lambda+\min(a(\ell),a(\bar\ell))$; summing proves~\eqref{eq:smallS}.
\end{proof}

\subsection{Comparison with the trace chain}
When $e_0\le\delta/2$, Lemma~\ref{lem:threshold} gives a set $S$ such that every assignment to its variables violates clauses contained in $S$ of total $\widehat{\pi}$-weight at least $\delta/2$. The lemma also gives the upper bound $\sum_{s\in S}\mu(s)a(s)\le2\lambda+B_{\widehat{\pi}}$. We seek a lower bound on the same sum. However, the walk using only internal clauses need not have a stationary distribution that is equal at complementary literals. We instead observe $P$ at successive visits to $S$. The resulting trace chain has the required stationary distribution and retains all original transitions within $S$, allowing Lemma~\ref{lem:oneway} to use the internal clause weights directly. We transfer the resulting affinity bound to the original walks of $P$ in Lemma~\ref{lem:comparison}.

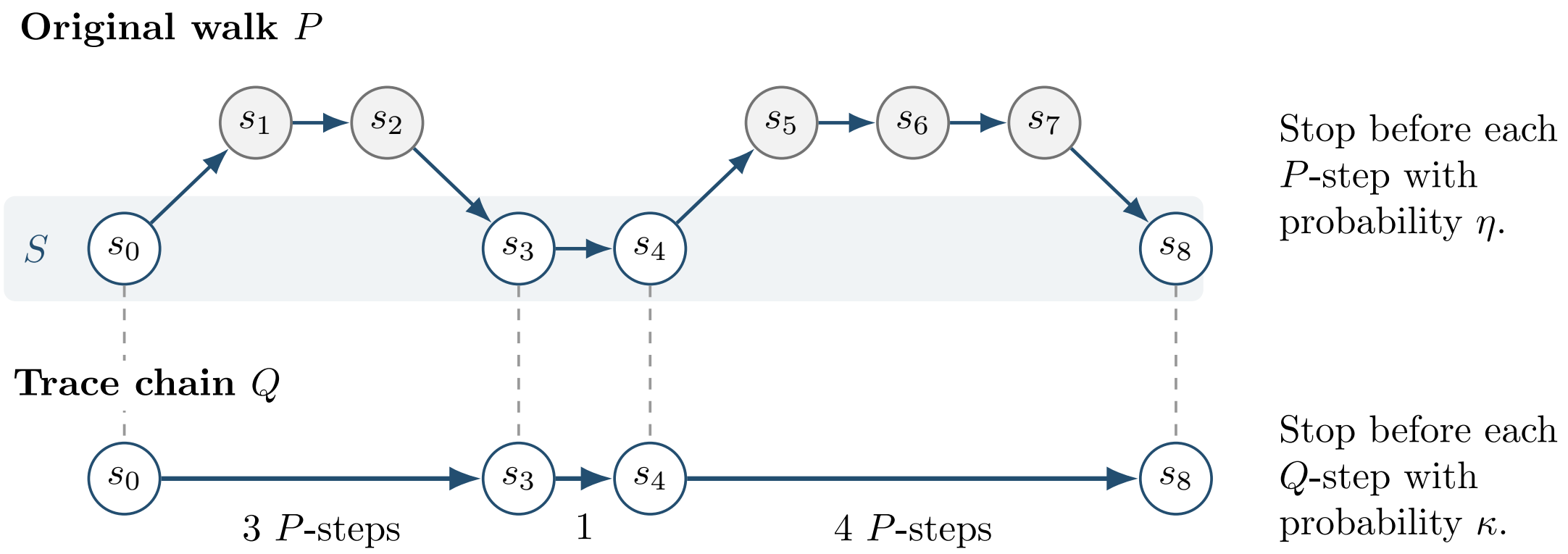
\begin{figure}[tbp]
 \centering
 \begin{tikzpicture}[manuscript diagram,
 visit/.style={literal node,minimum size=6.5mm},
 outside/.style={circle,draw=black!55,fill=black!5,inner sep=1pt,minimum size=6.5mm}]
 \node[anchor=west,font=\small\bfseries] at (-.4,2)
   {Original walk $P$};
 \fill[walkblue!7,rounded corners=3pt] (-.4,-.48) rectangle (10.55,.48);
 \node[anchor=east,text=walkblue] at (.15,0) {$S$};
 \foreach \name/\xx in {p0/.7,p3/4.3,p4/5.5,p8/10.3}
   \coordinate (\name) at (\xx,0);
 \node[visit] (x0) at (p0) {$s_0$};
 \node[visit] (x3) at (p3) {$s_3$};
 \node[visit] (x4) at (p4) {$s_4$};
 \node[visit] (x8) at (p8) {$s_8$};
 \foreach \ii/\xx in {1/1.9,2/3.1,5/6.7,6/7.9,7/9.1}
   \node[outside] (x\ii) at (\xx,1.15) {$s_{\ii}$};
 \foreach \from/\to in {0/1,1/2,2/3,3/4,4/5,5/6,6/7,7/8}
   \draw[diagram arrow] (x\from) -- (x\to);
 \node[align=left,text width=2.7cm,anchor=west] at (11.1,.65)
   {Stop before each\\$P$-step with\\probability $\eta$.};

 \foreach \ii/\xx in {0/.7,3/4.3,4/5.5,8/10.3} {
   \node[visit] (q\ii) at (\xx,-2.1) {$s_{\ii}$};
   \draw[dashed,black!40] (x\ii.south) -- (q\ii.north);
 }
 \node[anchor=west,font=\small\bfseries,fill=white,inner sep=2pt]
   at (-.4,-1.25) {Trace chain $Q$};
 \draw[diagram arrow,line width=1.2pt] (q0) -- (q3)
   node[midway,below=5pt] {$3$ $P$-steps};
 \draw[diagram arrow,line width=1.2pt] (q3) -- (q4)
   node[midway,below=5pt] {$1$};
 \draw[diagram arrow,line width=1.2pt] (q4) -- (q8)
   node[midway,below=5pt] {$4$ $P$-steps};
 \node[align=left,text width=2.7cm,anchor=west] at (11.1,-2.1)
   {Stop before each\\$Q$-step with\\probability $\kappa$.};
\end{tikzpicture}

 \caption{A path of $P$ and its successive visits to the complement-closed
 set $S$. The shaded band marks the visits to $S$ along this path.
 The three $Q$-steps expand to three, one, and four $P$-steps, respectively;
 the dashed vertical lines identify the corresponding visits.
 Geometric stopping with probability $\eta$ for $P$ acts before each original
 step, while stopping with probability $\kappa$ for $Q$ acts before each
 step of the trace chain. The two stopping rules therefore count different numbers
 of steps along the same expanded path.}
 \label{fig:return-chain}
\end{figure}

For the next two lemmas, take any complement-closed set $S\subseteq\mathcal L_+$ with $\mu(S)>0$. Write $\mu|_S:S\to[0,1]$ for the unnormalized restriction and let $\mu_S:S\to[0,1]$ be the probability mass function $\mu_S\coloneqq (\mu|_S)/\mu(S)$. Start $P$ at a state of $S$ and stop at its first return to $S$ after at least one step; let $Q\in[0,1]^{S\times S}$ be the transition kernel of the trace chain on $S$. Figure~\ref{fig:return-chain} illustrates how successive returns form a path of $Q$. Since the chain is finite and its stationary distribution $\mu$ is positive on $\mathcal L_+$, every state in $\mathcal L_+$ belongs to a recurrent class. A walk starting in $S$ stays in a recurrent class meeting $S$, so its next return to $S$ is almost surely finite. In formulas involving $Q$, $C$ and $I$ act on $\mathbb R^S$.

\begin{lemma}\label{lem:trace}
The trace kernel $Q$ has stationary distribution $\mu_S$, satisfies $Q(s,s)\ge1/2$, and obeys the entrywise inequality $2Q-I\ge T_{SS}$, where $T_{SS}$ is the restriction of $T$ to $S\times S$. Starting from $\mu_S$, the expected number of $P$-steps until the first return to $S$ is at most $1/\mu(S)$.
\end{lemma}
\begin{proof}
For $A,B\subseteq\mathcal L_+$, write $P_{AB}\in[0,1]^{A\times B}$ for the corresponding submatrix of $P$. Every one-step transition within $S$ is already a return, so $Q\ge P_{SS}$. In particular, $Q(s,s)\ge1/2$ and $2Q-I\ge2P_{SS}-I=T_{SS}$.

To prove $\mu_SQ=\mu_S$ and bound the expected number of $P$-steps per return, restrict to the recurrent classes meeting $S$ and let $U\subseteq\mathcal L_+\setminus S$ be their remaining states. A walk starting in $S$ cannot reach any other recurrent class. Every state in $U$ reaches $S$ almost surely. Since $U$ is finite, the series
\[
 (I-P_{UU})^{-1}=\sum_{r\ge0}P_{UU}^r
\]
converges, where $I$ is the identity on $\mathbb R^U$. Its $(s,z)$-entry is the expected number of visits to $z\in U$ before reaching $S$, starting at $s\in U$. Decomposing a return according to whether its first step stays in $S$ gives
\[
 Q=P_{SS}+P_{SU}(I-P_{UU})^{-1}P_{US}.
\]
Writing $\mu|_U:U\to[0,1]$ for the unnormalized restriction to $U$, stationarity gives
\[
 \mu|_U=(\mu|_S)P_{SU}(I-P_{UU})^{-1},\qquad \mu_SQ=\mu_S.
\]
Finally, the initial step contributes one to the return time, and the remaining steps are counted by the visits to $U$. Thus, with $\mathbf 1_U\in\mathbb R^U$ denoting the all-ones vector, the expected number of steps per return is
\[
 1+\mu_SP_{SU}(I-P_{UU})^{-1}\mathbf 1_U
 =1+\frac{\mu(U)}{\mu(S)}\le\frac{1}{\mu(S)}.\qedhere
\]
\end{proof}
The tester does not construct $Q$. In the analysis, each $Q$-transition expands into a path of $P$, whose non-lazy steps are original implications.

It remains to transfer affinity from $Q$ to $P$. Each $Q$-step can expand into a different number of $P$-steps. We weight each expanded path by its survival probability when stopping independently with probability $\eta$ before each $P$-step. In the proof below, we use the expected return-time bound from Lemma~\ref{lem:trace} to bound the mass lost under this weighting and compare the surviving path weights with those of $R_\eta$.
For a stopping probability $0<\kappa<1$, use $R^Q_\kappa\in[0,1]^{S\times S}$ and $a^Q_\kappa:S\to[0,1]$ from~\eqref{eq:geometrickernel} with kernel $Q$ on $S$. Thus $R^Q_\kappa(s,\cdot)$ is the endpoint distribution when $Q$ starts at $s$ and stops with probability $\kappa$ before each trace step.

\begin{lemma}[Comparison with the trace chain]\label{lem:comparison}
Suppose $\mu(S)\ge q>0$ and $0<\kappa\le1/2$, and assume
\[
 \E_{s\sim\mu_S}a^Q_\kappa(s)\ge h>0.
\]
Choose
\[
 r_0\coloneqq qh^2/16,\qquad \eta\coloneqq r_0\kappa.
\]
Then the endpoint kernel $R_\eta$ satisfies
\begin{equation}\label{eq:transfer}
 \sum_{s\in S}\mu(s)
 \Aff(R_\eta(s,\cdot),CR_\eta(s,\cdot))
 \ge \frac{q r_0h}{2}.
\end{equation}
\end{lemma}
\begin{proof}
Start the $Q$-walk from $\mu_S$ and take an independent number $J\in\mathbb Z_{\ge0}$ of steps with $\Pr[J=j]=\kappa(1-\kappa)^j$. Realize each $Q$-step by running $P$ from its current state until its next return to $S$, and let $K\in\mathbb Z_{\ge0}$ be the total number of $P$-steps. Every $Q$-step starts with distribution $\mu_S$, so Lemma~\ref{lem:trace} gives $\E K\le1/(\kappa\mu(S))$.

Retain each expanded path with probability $(1-\eta)^K$. Let $G:S\times\mathcal L_+\to[0,1]$ be the resulting joint subprobability mass function of seed and endpoint, zero outside $S\times S$, and define $G^C:S\times\mathcal L_+\to[0,1]$ by $G^C(s,z)\coloneqq G(s,\bar z)$. The removed mass $\epsilon\in[0,1]$ satisfies
\[
 \epsilon=\E\bigl[1-(1-\eta)^K\bigr]
 \le\eta\E K\le\frac{r_0}{\mu(S)}\le\frac{h^2}{16}.
\]
Before retention, the joint distributions $\mu_S(s)R^Q_\kappa(s,z)$ and $\mu_S(s)R^Q_\kappa(s,\bar z)$ have affinity $\E_{s\sim\mu_S}a^Q_\kappa(s)\ge h$. Reducing each of two probability measures by mass at most $\epsilon$ decreases their affinity by at most $2\sqrt\epsilon$, as follows from Cauchy--Schwarz applied to the differences of their square roots. Hence
\[
 \Aff(G,G^C)\ge h-2\sqrt\epsilon\ge h/2.
\]

To transfer this affinity bound to $R_\eta$, we now compare the retained path weights with the probabilities of the same paths under $R_\eta$.
A path $(s_0,\ldots,s_r)\in\mathcal L_+^{r+1}$ of $P$ with $s_0,s_r\in S$ has a unique decomposition into first-return paths to $S$. If it contains $j$ returns, its retained weight, conditional on $s_0$, is
\[
 \kappa(1-\kappa)^j(1-\eta)^r\prod_{i=0}^{r-1}P(s_i,s_{i+1}).
\]
Under the stopping rule for $R_\eta$, the probability of $(s_0,\ldots,s_r)$, conditional on $s_0$, is
\[
 \eta(1-\eta)^r\prod_{i=0}^{r-1}P(s_i,s_{i+1}).
\]
Since $\eta=r_0\kappa$ and $(1-\kappa)^j\le1$, the $R_\eta$ path probability is at least $r_0$ times the retained weight. Summing over paths gives
\begin{equation}\label{eq:trace-endpoint-domination}
 \mu_S(s)R_\eta(s,z)\ge r_0G(s,z).
\end{equation}
Applying~\eqref{eq:trace-endpoint-domination} at $z$ and $\bar z$, taking square roots, and summing, we obtain
\[
 \E_{s\sim\mu_S}\Aff(R_\eta(s,\cdot),CR_\eta(s,\cdot))
 \ge r_0\Aff(G,G^C)\ge r_0h/2.
\]
Multiplying by $\mu(S)\ge q$ proves~\eqref{eq:transfer}.
\end{proof}
The tester samples $R_\eta$ directly, without detecting $S$ or its return times.

We now choose the parameters so that the upper bound~\eqref{eq:smallS} and the lower bound~\eqref{eq:transfer} on $\sum_{s\in S}\mu(s)a(s)$ imply $B_{\widehat{\pi}}\ge2\lambda$.

\begin{proof}[Proof of Theorem~\ref{thm:geo}]
Fix $0<\delta\le1$. In Lemma~\ref{lem:threshold}, the total $\widehat{\pi}$-weight of clauses with exactly one endpoint in $S$ and violated clauses with both endpoints outside $S$ is at most $e_0=\sqrt D/\lambda$. We will choose the parameters so that $e_0\le\delta/2$. Then $\mu(S)\ge\delta/2$, and every assignment to the variables of $S$ violates clauses contained in $S$ of total $\widehat{\pi}$-weight at least $\delta/2$. For the trace chain, Lemma~\ref{lem:oneway} will then give $\E_{s\sim\mu_S}a^Q_\kappa(s)\ge\delta^2/8$ when its stopping probability is small enough. Use these lower bounds on $\mu(S)$ and $\E_{s\sim\mu_S}a^Q_\kappa(s)$ in Lemma~\ref{lem:comparison}:
\[
 q\coloneqq \delta/2,\qquad h\coloneqq \delta^2/8,\qquad r_0\coloneqq qh^2/16.
\]
The transferred lower bound in~\eqref{eq:transfer} is $qr_0h/2$. To make this bound equal to $4\lambda$, choose
\[
 \lambda\coloneqq qr_0h/8,\qquad b_\delta\coloneqq 2\lambda.
\]
The constants $q,h,r_0,\lambda,b_\delta$ depend only on $\delta$, and $0<\lambda<1/2$.

Choose the stopping probabilities by
\[
 \eta\coloneqq \min\left\{\frac{r_0\delta^2}{16\log_2(2n)},
                         \frac{\delta^2\lambda^2}{512}\right\},
 \qquad \kappa\coloneqq \eta/r_0.
\]
The first term ensures $\kappa\le\delta^2/(16\log_2(2n))$, which will supply the affinity bound for the trace chain. Both $\eta$ and $\kappa$ lie in $(0,1/2]$ and are determined by $\delta,n$ alone, before we choose $S$. The second term and Lemma~\ref{lem:stability} give
\[
 D\le32\eta\le\delta^2\lambda^2/16.
\]
Thus $e_0=\sqrt D/\lambda\le\delta/4<\delta/2$ in Lemma~\ref{lem:threshold}.
Choose the set $S$ supplied by Lemma~\ref{lem:threshold}. Then $\mu(S)\ge q$, and every assignment on the variables of $S$ violates internal clauses of total $\widehat{\pi}$-weight at least $\delta/2$.

To apply Lemma~\ref{lem:oneway} to $Q$, divide the original internal clause weights by $\mu(S)$, matching the normalization of $\mu_S$. Define $\omega\in\mathbb R_{\ge0}^m$ by $\omega_e\coloneqq\widehat{\pi}_e/\mu(S)$ for $e\subseteq S$ and $\omega_e\coloneqq0$ otherwise. Then $\tau_\omega(F)\ge\delta/(2\mu(S))\ge\delta/2$. By Lemma~\ref{lem:trace}, $Q$ is lazy with stationary distribution $\mu_S$, and for $i,j\in S$,
\[
 \sum_{e=\{i,\bar j\}}\omega_e
 =2\mu_S(i)T(i,j)
 \le2\mu_S(i)(2Q-I)(i,j).
\]
Since $\mu_S(i)=\mu_S(\bar i)$, Lemma~\ref{lem:oneway} applies with $\Omega=S$, $K=Q$, $\nu=\mu_S$, and the stopping probability $\kappa$. It gives
\[
 \E_{s\sim\mu_S}a^Q_\kappa(s)
 \ge\frac{\delta^2}{4}-2\kappa\log_2(2n)
 \ge\frac{\delta^2}{8}=h.
\]

Lemma~\ref{lem:comparison} therefore gives $\sum_{s\in S}\mu(s)a(s)\ge qr_0h/2=4\lambda$. Equation~\eqref{eq:smallS} bounds this same sum above by $2\lambda+B_{\widehat{\pi}}$. Hence
\[
 2\lambda+B_{\widehat{\pi}}\ge4\lambda,
 \qquad\text{so}\qquad B_{\widehat{\pi}}\ge2\lambda=b_\delta.
\]
It remains to bound the expected walk length. Substituting $r_0=2^{-11}\delta^5$ and $\lambda=2^{-18}\delta^8$ into the choice of $\eta$ gives
\begin{equation}\label{eq:geostopping}
 \eta^{-1}=\max\left\{2^{15}\delta^{-7}\log_2(2n),\,
                              2^{45}\delta^{-18}\right\}.
\end{equation}
Since $\log_2(2n)\le2\log(2n)$ and $\log(2n)\ge1/2$, the following explicit constants satisfy the theorem:
\begin{equation}\label{eq:geoc}
 b_\delta=2^{-17}\delta^8,\qquad C_\delta\coloneqq 2^{46}\delta^{-18}.
\end{equation}
In particular, $\eta^{-1}=O(\delta^{-18}+\delta^{-7}\log(2n))$, with absolute implicit constants.
\end{proof}

\section{The Tester and Its Analysis}\label{sec:tester}
We combine the preceding sections to construct the tester and prove Theorem~\ref{thm:tester}. Section~\ref{sec:geo} gives a lower bound on the two-way affinity $B_{\widehat{\pi}}$ for the balanced comparison distribution $\widehat{\pi}$, whereas the tester uses the locally computed weights $w$ from Section~\ref{sec:localweights}, normalized to $\pi=w/W(w)$. We first transfer that bound to the tester's walks using the relative-entropy guarantees of Section~\ref{sec:comparison}, then show how endpoint collisions yield contradictory paths. We next give the E2SAT algorithm and bound its query cost and error probability, and finally reduce general 2SAT to E2SAT while preserving the ability to output a contradiction in the original clauses.

\subsection{Two-way affinity transfer}\label{sec:transfer}
Our goal is to turn the lower bound on $B_{\widehat{\pi}}$ into a lower bound on $B_\pi$. We first bound the relative entropy between the joint seed and endpoint distributions, then use it to control the loss in two-way affinity. The seeds are drawn from $\mu_{\widehat{\pi}}$ and $\mu_\pi$, respectively; stationarity is required only for the former.

We use the walk and affinity notation of Section~\ref{sec:geo} for the clause distributions $\widehat{\pi}$ and $\pi$, with a common stopping probability $\eta$. All kernels here act on the full literal set $\mathcal L$, including literals of zero incident weight. We suppress $\eta$ in the affinity notation, writing $a_\sigma(s)$ and $B_\sigma$ for $\sigma\in\{\widehat{\pi},\pi\}$.

\begin{lemma}[Relative entropy of walk endpoints]\label{lem:walkkl}
Let $\widehat{\pi},\pi\in[0,1]^m$ be clause probability distributions, with $\widehat{\pi}$ balanced and supported where $\pi$ is positive. When evaluating $F_\pi(\widehat{\pi})$, omit clauses with $\pi_e=0$, which also have $\widehat{\pi}_e=0$ by the support assumption; this extends~\eqref{eq:F} to the present setting. Define the nonnegative scalars $K_0,K_1$ by $K_0\coloneqq \KL(\mu_{\widehat{\pi}}\|\mu_\pi)$ and $K_1\coloneqq F_\pi(\widehat{\pi})$. In the $\widehat{\pi}$ experiment, draw the seed $s\sim\mu_{\widehat{\pi}}$; in the $\pi$ experiment, draw $s\sim\mu_\pi$. Start the walk either at $s$ or at $\bar s$, and report either the endpoint or its complement, making the same two choices in both experiments. The joint distribution of the seed label $s$ and reported endpoint under $\widehat{\pi}$ has divergence from its counterpart under $\pi$ at most
\begin{equation}\label{eq:walkdivergence}
 d\coloneqq K_0+\frac{1-\eta}{2\eta}K_1\le K_0+\frac{K_1}{2\eta}.
\end{equation}
\end{lemma}
\begin{proof}
Retain the seed, geometric length, lazy-step indicators, and chosen clauses. The relative entropy between the seed distributions is $K_0$; the geometric length and lazy-step indicators have the same distributions in the $\widehat{\pi}$ and $\pi$ experiments. Under $\widehat{\pi}$, the starting state has distribution $\mu_{\widehat{\pi}}$ even when complemented, and every subsequent state has distribution $\mu_{\widehat{\pi}}$ by stationarity. By~\eqref{eq:conditional}, at each non-lazy step the relative entropy between $q_{\widehat{\pi}}(\cdot\mid\ell)$ and $q_\pi(\cdot\mid\ell)$, averaged over $\ell\sim\mu_{\widehat{\pi}}$, is $K_1$. The expected number of non-lazy steps is $(1-\eta)/(2\eta)$. The chain rule gives divergence $K_0+(1-\eta)K_1/(2\eta)$ for the complete transcript. Discarding all but the seed and reported endpoint cannot increase divergence, which proves~\eqref{eq:walkdivergence}. Complementing the endpoint is a bijection and does not change divergence.
\end{proof}

We next use the divergence bound in Lemma~\ref{lem:walkkl} to control the loss in two-way affinity.
\begin{lemma}[Comparing endpoint affinities]\label{lem:kltransfer}
Under the assumptions of Lemma~\ref{lem:walkkl}, let $d\in\mathbb R_{\ge0}$ be as in~\eqref{eq:walkdivergence}. If $B_{\widehat{\pi}}\ge b>0$, then
\begin{equation}\label{eq:kltransfer}
 B_\pi\ge\frac{b}{2}\exp\left(-\frac{64(d+1)}{b^2}\right).
\end{equation}
\end{lemma}
\begin{proof}
For $j\in\{0,1\}$ and $s\in\mathcal L$, put $s_0\coloneqq s$ and $s_1\coloneqq \bar s$. Define the joint probability mass functions $Q_j,P_j:\mathcal L\times\mathcal L\to[0,1]$ by
\[
 Q_j(s,z)\coloneqq \mu_{\widehat{\pi}}(s)R_{\widehat{\pi},\eta}(s_j,z),\qquad
 P_j(s,z)\coloneqq \mu_\pi(s)R_{\pi,\eta}(s_j,z).
\]
Lemma~\ref{lem:walkkl} gives $\KL(Q_j\|P_j)\le d$. In our application, $d$ may grow with $\log n$, so we seek a multiplicative comparison that can tolerate this growth. For a parameter $L>0$, if $Q_j\le e^LP_j$ pointwise, the affinity between $P_j(s,z)$ and $P_j(s,\bar z)$ is at least $e^{-L}$ times the corresponding affinity for $Q_j$. We enforce this inequality by truncating $Q_j$ where its ratio to $P_j$ is too large. Relative entropy controls the mass removed, so enough affinity survives the truncation. We will choose $L$ proportional to $(d+1)/b^2$; the factor $e^{-L}$ will then give the exponential dependence on $d$ in~\eqref{eq:kltransfer}.
To bound the mass removed by truncation, let $Q,P$ be probability mass functions on a common finite set with $\KL(Q\|P)\le d$. At a point with $0<Q<P$, put $t\coloneqq Q/P$. The inequality $t\log(1/t)\le1/e$ gives
\[
 Q\log(P/Q)=Pt\log(1/t)\le P/e.
\]
Thus the $Q$-expectation of the negative part of $\log(Q/P)$ is at most $\sum_{Q<P}P/e\le1/e$. The positive part of $\log(Q/P)$ therefore has expectation at most $d+1$. Markov's inequality gives $Q\{Q>e^LP\}\le(d+1)/L$ for every $L>0$.

Accordingly, define the subprobability mass functions $\widehat Q_j:\mathcal L\times\mathcal L\to[0,1]$ and their removed masses $\delta_j\in[0,1]$ by
\[
 \widehat Q_j(s,z)\coloneqq \min\{Q_j(s,z),e^LP_j(s,z)\},\qquad
 \delta_j\coloneqq \sum_{s,z}(Q_j-\widehat Q_j)(s,z)\le\frac{d+1}{L}.
\]
Define the affinity contributions $A_j,\widehat A_j:\mathcal L\to[0,1]$ at each seed, including the seed mass, by
\[
 \begin{aligned}
 A_j(s)&\coloneqq \sum_z\sqrt{Q_j(s,z)Q_j(s,\bar z)}
       =\mu_{\widehat{\pi}}(s)a_{\widehat{\pi}}(s_j),\\
 \widehat A_j(s)&\coloneqq \sum_z\sqrt{\widehat Q_j(s,z)\widehat Q_j(s,\bar z)}.
 \end{aligned}
\]
Using the Euclidean norm over seed--endpoint pairs, we have $\|\sqrt{Q_j}-\sqrt{\widehat Q_j}\|_2^2\le\delta_j$. Cauchy--Schwarz therefore bounds the total affinity loss by
\[
 \sum_s\bigl(A_j(s)-\widehat A_j(s)\bigr)
 \le2\|\sqrt{Q_j}-\sqrt{\widehat Q_j}\|_2
 \le2\sqrt{\delta_j}.
\]
Also, $\widehat A_j(s)\le e^L\mu_\pi(s)a_\pi(s_j)$. We need both affinity contributions $\widehat A_0(s)$ and $\widehat A_1(s)$ at the same seed: taking the minimum over $j$ before summing over $s$ loses at most the sum of the affinity decreases. Consequently,
\[
 e^LB_\pi\ge\sum_s\min_j\widehat A_j(s)
 \ge B_{\widehat{\pi}}-2\sqrt{\delta_0}-2\sqrt{\delta_1}
 \ge B_{\widehat{\pi}}-4\sqrt{\frac{d+1}{L}}.
\]
Taking $L\coloneqq 64(d+1)/b^2$ and using $B_{\widehat{\pi}}\ge b$ proves~\eqref{eq:kltransfer}.
\end{proof}
In the proof of Lemma~\ref{lem:kltransfer}, we apply the divergence bound $d$ to the two pairs $(Q_j,P_j)$, $j=0,1$, of joint seed and endpoint distributions. Thus the divergence bound $d$ in~\eqref{eq:walkdivergence} is not multiplied by the number of walks that the tester later samples.

\subsection{Endpoint collisions}\label{sec:collision}
We now use $B_\pi$ from~\eqref{eq:distributionaffinity} to bound the number of walks needed to find a contradiction. The next lemma bounds the probability that independent samples from two distributions contain a common element in terms of their Hellinger affinity. We will apply it to an endpoint distribution and its complement.
\begin{lemma}[Two-distribution collision bound]\label{lem:collision}
There is a universal constant $C_*$ with the following property. Let $p,q:\Omega\to[0,1]$ be probability mass functions on a finite set $\Omega$ of size at most $N$, with Hellinger affinity $\Aff(p,q)\ge a>0$. Two independent batches of samples, one from $p$ and one from $q$, contain a common element with probability at least $1/4$ if each batch has size
\[
 M\coloneqq \left\lceil C_*\left(\frac{\sqrt N}{a}+\frac{1}{a^2}\right)\right\rceil.
\]
\end{lemma}
\begin{proof}
For completeness, we prove the collision bound of Yoshida~\cite[Lemma~2.1]{manuscript} by a second-moment argument. Let $X_1,\ldots,X_M\in\Omega$ and $Y_1,\ldots,Y_M\in\Omega$ be the independent batches drawn from $p$ and $q$, respectively, and let $N_i,K_i\in\{0,\ldots,M\}$ count the occurrences of $i\in\Omega$ in the two batches. Put $I\coloneqq \{i\in\Omega:p_iq_i>0\}$ and $A\coloneqq \Aff(p,q)\ge a$. The normalized collision count $M^{-2}\sum_iN_iK_i$ has expectation $\sum_i p_iq_i$. To make this expectation equal the affinity $A=\sum_i\sqrt{p_iq_i}$, weight each collision at $i\in I$ by $1/\sqrt{p_iq_i}$. This changes the value of the count but preserves whether any collision occurred. Accordingly, define the nonnegative random variable
\[
 Z\coloneqq \frac{1}{M^2}\sum_{i\in I}\frac{N_iK_i}{\sqrt{p_iq_i}}.
\]
Then $Z>0$ exactly when the batches collide, and $\E Z=A$. Expanding $\E Z^2$ according to whether two sample pairs $(X_r,Y_t)$ and $(X_{r'},Y_{t'})$ share neither, one, or both indices gives
\[
 \E Z^2
 =\frac{(M-1)^2A^2+(M-1)\sum_{i\in I}(p_i+q_i)+|I|}{M^2}
 \le A^2+\frac{2}M+\frac{N}{M^2}.
\]
Therefore Cauchy--Schwarz yields
\[
 \Pr[Z>0]\ge\frac{(\E Z)^2}{\E Z^2}
 \ge\frac{A^2}{A^2+2/M+N/M^2}\ge\frac{1}{4}
\]
when $M\ge\sqrt N/a+a^{-2}$. Thus we may take $C_*\coloneqq 1$. The weights in $Z$ are used only in the analysis; the tester simply checks for a common sample.
\end{proof}

To apply Lemma~\ref{lem:collision} to the contradiction search, suppose $B_\pi\ge b_0$ for some $b_0\in(0,1]$. Then $\min\{a_\pi(s),a_\pi(\bar s)\}\ge b_0/2$ on seeds of total $\mu_\pi$-mass at least $b_0/2$, because this minimum lies in $[0,1]$. For such a seed $s$, use four independent batches of forward walks: two from $s$ and two from $\bar s$. Within each pair, seek complementary endpoints. Lemma~\ref{lem:collision} with $a=b_0/2$ gives paths $s\leadsto\bar s$ and $\bar s\leadsto s$ with probability at least $1/16$. A trial that draws $s\sim\mu_\pi$ and performs these four batches therefore succeeds with probability at least $b_0/32$. Taking
\begin{equation}\label{eq:RM}
 R\coloneqq \left\lceil\frac{128}{b_0}\right\rceil,\qquad
 M\coloneqq \left\lceil C_*\left(\frac{2\sqrt{2n}}{b_0}+\frac{4}{b_0^2}\right)\right\rceil
\end{equation}
as the number of independent seed trials and the number of endpoints per batch, respectively, makes the probability of finding no contradiction at most $1/16$. The total number of endpoints is
\begin{equation}\label{eq:Q}
 Q\coloneqq 4RM=O(\sqrt n/b_0^2+b_0^{-3}).
\end{equation}
Each sampled path is retained with the clauses used by its non-lazy steps. If both endpoint tests succeed, reverse and complement one path in each pair and concatenate the paths to obtain $s\leadsto\bar s\leadsto s$, as in Figure~\ref{fig:collision-certificate}.

\subsection{The E2SAT tester}\label{sec:algorithm}
The tester evaluates clause weights on demand, samples seeds and walks using those weights, and searches for contradictory paths. We give the procedure first, then choose its parameters and specify how to sample within deterministic budgets.

Write $D$ for the E2SAT degree bound, and fix $0<\nu<1/2$. The far condition is $\tau(F)\ge\varepsilon Dn$, and $m\le Dn/2$ gives $\tau(F)/m\ge2\varepsilon$ on a far input.

\paragraph{The algorithm.}
Use the local oracle of Theorem~\ref{thm:local} to evaluate a deterministic clause-weight vector $w\in(0,1]^m$ as needed. The procedure uses a target lower bound $b_0\in(0,1]$ on the two-way affinity $B_\pi$ and a stopping probability $\eta\in(0,1)$. Set the number of seed trials $R$, the batch size $M$, and the total number of walks $Q=4RM$ by~\eqref{eq:RM}--\eqref{eq:Q}. The oracle parameters, $b_0$, $\eta$, and the budgets below depend only on $n,\varepsilon,D,\nu$ and are specified in the following paragraphs.

For each of the $R$ trials, attempt to sample a seed using the incidence-slot procedure below. As we verify below, the accepted seed has distribution $\mu_\pi$, where $\pi\coloneqq w/W(w)$. Run four independent batches of $M$ forward lazy walks with weights $w$, two from the seed and two from its complement. Each walk stops independently with probability $\eta$ before each step. Test each pair of batches for complementary endpoints, retaining the paths and the clauses used by their non-lazy steps. If both tests succeed for the same seed, reject and output the resulting contradictory paths. Otherwise accept after the $R$ trials. Also accept immediately if the total number of seed proposals reaches its budget $S_{\max}$ without obtaining all seeds, or if the total number of walk steps reaches its budget $L_{\max}$ before all walks finish.

\paragraph{Parameter choices and two-way affinity guarantee.}
Choose $\delta_0>0$ so that $\tau(F)\ge\delta_0m$ on far inputs, as required by the local oracle. Let $k>0$ be the coefficient in its total-weight bound $W(w)\ge km$, and let $g,h>0$ be the lower bounds we seek for $\tau_\pi(F)$ and $\tau_{\widehat{\pi}}(F)$, respectively:
\begin{equation}\label{eq:gapparameters}
 \delta_0\coloneqq \min\{\varepsilon,1/4\},\qquad
 k\coloneqq k_{\delta_0}=\frac{\delta_0}{4}e^{-4/\delta_0},\qquad
 g\coloneqq k/2,\qquad h\coloneqq g^2/1600.
\end{equation}
Theorem~\ref{thm:geo}, applied with $\delta=h$, supplies a lower bound $B_{\widehat{\pi}}\ge b_h$ and a stopping probability with $\eta^{-1}\le C_h\log(2n)$. Use the constants $b_h,C_h$ from~\eqref{eq:geoc} and write $b\coloneqq b_h$. Choose $\alpha$ to make the divergence accumulated over $O(\log n)$ steps contribute an exponent at most $\nu/2$ in the lower bound on $B_\pi$ from Lemma~\ref{lem:kltransfer}. Then choose the imbalance tolerance $\beta$ to meet Theorem~\ref{thm:comparison}, and the number $T_0$ of descent iterates to meet that tolerance:
\begin{equation}\label{eq:params}
 \alpha\coloneqq \min\left\{1,\frac{\nu b^2}{32C_h}\right\},\qquad
 \beta\coloneqq \min\{1,\alpha g^2/400\},\qquad
 T_0\coloneqq \left\lceil\frac{288D}{k^2\beta^2}\right\rceil.
\end{equation}
The constants in~\eqref{eq:gapparameters} and~\eqref{eq:params} depend only on $\varepsilon,D,\nu$. Invoke Theorem~\ref{thm:local} with $\delta=\delta_0$, degree bound $D$, and imbalance tolerance $\beta$. The oracle locally evaluates the deterministic average $w=T_0^{-1}\sum_{j=0}^{T_0-1}w(z^j)\in(0,1]^m$ from~\eqref{eq:averagedweights}.

Put $W\coloneqq W(w)$. On every far input, Theorem~\ref{thm:local} gives
\begin{equation}\label{eq:weightbounds}
 W\ge km,\qquad \|b(w)\|_1/W\le\beta,\qquad
 \tau_\pi(F)\ge g,\quad \pi_e=w_e/W.
\end{equation}
Every clause has positive weight, as required in Section~\ref{sec:comparison}. The tester need not check~\eqref{eq:weightbounds} or compute $m,W$, or $\pi$ explicitly.

On a far input, applying Theorem~\ref{thm:comparison} to~\eqref{eq:weightbounds}, with $\delta=g$, provides a balanced distribution $\widehat{\pi}$ with $\tau_{\widehat{\pi}}(F)\ge h$, $K_0=\KL(\mu_{\widehat{\pi}}\|\mu_\pi)\le\log(1/(2h))$, and $K_1=F_\pi(\widehat{\pi})\le\alpha/2$. Choose the stopping probability $\eta=\eta(h,n)$ by~\eqref{eq:geostopping} with $\delta=h$. Theorem~\ref{thm:geo} gives $B_{\widehat{\pi}}\ge b$, while Lemmas~\ref{lem:walkkl} and~\ref{lem:kltransfer} give the following bound with constants $c,\gamma>0$:
\begin{align}
 B_\pi
 &\ge c\,(2n)^{-\gamma},\notag\\
 c&\coloneqq \frac{b}{2}\exp\left[-\frac{64(1+\log(1/(2h)))}{b^2}\right],\qquad
 \gamma\coloneqq \frac{16\alpha C_h}{b^2}\le\nu/2.\label{eq:actualoverlap}
\end{align}
Since $\gamma\le\nu/2$, we may use the following lower bound on $B_\pi$:
\begin{equation}\label{eq:bzero}
 b_0\coloneqq c(2n)^{-\nu/2}.
\end{equation}
Then $B_\pi\ge b_0$ and $b_0^{-1}=O_{\varepsilon,D,\nu}(n^{\nu/2})$.

\paragraph{Exact seed sampling from incidence slots.}
Choose a uniform variable and a uniform incidence slot in $\{1,\ldots,D\}$. If the answer is null, discard the proposal and try again, subject to the proposal budget. For a returned E2SAT clause $e$, compute $w_e$ and accept this proposal with probability $w_e$. The seed is the complement of the literal of the queried variable in the returned clause.

The vector $w$ is deterministic. Each incidence of a clause $e$ at one of its variables is proposed and accepted with probability $w_e/(nD)$. Thus accepted seeds have distribution $\mu_\pi$ exactly, with the associated clause distributed as $\pi$. Without a proposal budget, the accepted seeds are i.i.d. and independent of the acceptance indicators. The number of proposals needed to obtain $R$ accepted seeds depends only on those indicators, so imposing the budget can reduce the rejection probability only by the probability of exhausting it.

On a far input, we have $m\ge\tau(F)\ge\varepsilon Dn$. Together with the total-weight bound in~\eqref{eq:weightbounds}, this gives the following lower bound on the probability of accepting one seed proposal:
\[
 p\coloneqq \frac{2W}{nD}\ge\frac{2km}{nD}\ge p_0,\qquad p_0\coloneqq 2k\varepsilon.
\]
Cap all seed proposals at $S_{\max}\coloneqq \lceil64R/p_0\rceil$. Without this budget, sampling $R$ seeds on a far input takes at most $R/p_0$ proposals in expectation. By Markov's inequality, the probability of exhausting the proposal budget is at most $1/64$.

\paragraph{Walk steps and stopping.}
At a literal $\ell$, query the at most $D$ incidences of its variable and compute $w_e$ for each clause $e$ in which $\bar\ell$ appears. A non-lazy step uses these weights to sample $T_\pi(\ell,\cdot)$, as defined in Section~\ref{sec:geo}. The deterministic oracle gives consistent weights on repeated queries.

Each of the $Q$ walks has expected length $(1-\eta)/\eta\le1/\eta$, regardless of the seed and input. Cap the total number of steps taken by these walks at $L_{\max}\coloneqq \lceil64Q/\eta\rceil$. By Markov's inequality, the probability of exhausting this step budget is at most $1/64$.

\paragraph{Soundness and completeness.}
On a far input, we have $B_\pi\ge b_0$ by~\eqref{eq:actualoverlap}--\eqref{eq:bzero}. Thus, if no budget is imposed on seed proposals or walk steps, the probability that the tester finds no contradiction is at most $1/16$. Exhaustion of the seed proposal budget $S_{\max}$ or the step budget $L_{\max}$ adds at most $2/64$ to the failure probability. The tester therefore rejects with probability at least $1-1/16-2/64=29/32>2/3$.

The tester rejects only if it finds contradictory implication paths in the input. Thus the tester accepts every satisfiable input, regardless of the weights, random choices, or budget limits.

\paragraph{Query complexity.}
The local oracle requires $O(D^{T_0+2})$ incidence queries per clause weight. Since $T_0$ in~\eqref{eq:params} is independent of $n$, this cost depends only on $\varepsilon,D,\nu$. Thus each seed proposal and walk step uses $O_{\varepsilon,D,\nu}(1)$ queries. For the number of endpoints, substituting~\eqref{eq:bzero} into~\eqref{eq:Q} gives
\[
 Q=O_{\varepsilon,D,\nu}\left(n^{1/2+\nu}+n^{3\nu/2}\right)
 =O_{\varepsilon,D,\nu}(n^{1/2+\nu}).
\]
Seed proposals cost $O_{\varepsilon,D,\nu}(R)$ queries, and the step budget is $O_{\varepsilon,D,\nu}(Q\log(2n))$. Hence the query complexity is $\widetilde O_{\varepsilon,D,\nu}(n^{1/2+\nu})$. This proves Theorem~\ref{thm:tester} for E2SAT.

\subsection{Reduction from general 2SAT}\label{sec:general}
We complete the proof of Theorem~\ref{thm:tester} by extending the E2SAT tester to formulas with unit clauses and tautologies. The reduction must preserve the distance from satisfiability up to a constant factor, allow local simulation of queries, and convert a detected contradiction back to original clauses.

We construct an E2SAT formula $F'$ with $\tau(F')=\tau(F)$ and simulate each incidence query to $F'$ using one query to $F$. Normalize $(a\vee a)$ to the unit clause $(a)$ and discard tautologies $(a\vee\bar a)$, where $a$ is an original literal. Keep every clause on two distinct variables. For each original variable $x$, introduce a fresh Boolean variable $z_x$ and replace each unit clause on a literal $a\in\{x,\bar x\}$ by
\[
 (a)\quad\longmapsto\quad(a\vee z_x),(a\vee\bar z_x).
\]
The unit clauses $(x)$ and $(\bar x)$, when present, share $z_x$. Since the auxiliary variables are fresh, this construction introduces no repeated clauses. If $a$ is true, both replacement clauses are satisfied; if $a$ is false, exactly one is violated, regardless of $z_x$. Thus every assignment to the original and auxiliary variables violates exactly as many clauses of $F'$ as its restriction violates in $F$. In particular, $F'$ is satisfiable exactly when $F$ is, and $\tau(F')=\tau(F)$.

Include $z_x$ even when $x$ has no unit clauses, so that $F'$ has the known size $n'\coloneqq 2n$. Each original incidence produces at most two incidences at its original variable, and only a unit on $x$ produces incidences at $z_x$. Hence the maximum degree is at most $D\coloneqq 2\Delta$. To simulate the incidence oracle, assign output slots $2i-1,2i$ at both $x$ and $z_x$ to the original slot $(x,i)$:
\begin{itemize}
\item If $(x,i)$ contains a unit $(a)$ after normalization, slots $2i-1,2i$ at both $x$ and $z_x$ contain $(a\vee z_x),(a\vee\bar z_x)$, respectively.
\item If $(x,i)$ contains a clause on two distinct variables, slot $2i-1$ at $x$ contains that clause; slot $2i$ at $x$ and both slots at $z_x$ are null.
\item If $(x,i)$ is null or contains a tautology, all four output slots are null.
\end{itemize}
For each clause returned by the simulated oracle, retain the original clause that produced it, so that a contradiction in $F'$ can be converted to one in $F$. Every output query uses one original query. Sampling uniformly from the $2n$ labels $x,z_x$ requires no preprocessing or enumeration of clauses.

To verify preservation of distance, suppose the original formula is $\varepsilon$-far. Then
\[
 \tau(F')=\tau(F)\ge\varepsilon\Delta n
 =\varepsilon'Dn',\qquad
 \varepsilon'\coloneqq \varepsilon/4.
\]
We may therefore apply the E2SAT tester with parameters $D,\varepsilon',\nu$. Since $n'=2n$, the query bound in the original model remains
\[
 \widetilde O_{\varepsilon,\Delta,\nu}(n^{1/2+\nu}).
\]

Finally, suppose the tester finds a contradiction in $F'$. Let $F_0$ be the subformula of $F$ consisting of the original clauses that give rise to the clauses on the contradictory paths. The formula $F_0$ is unsatisfiable: any satisfying assignment would satisfy the retained binary clauses and both replacements of every unit, for any choice of auxiliary values, contradicting the paths in $F'$. Using the implication graph of $F_0$, with each unit clause $(a)$ represented by the implication $\bar a\to a$, the tester can extract paths $s\leadsto\bar s\leadsto s$ for some literal $s$ of $F_0$ without further queries. This proves Theorem~\ref{thm:tester} for general 2SAT.

\bibliographystyle{plain}

\end{document}